\documentclass[fleqn]{2017SCGE}
\usepackage{graphicx}
\usepackage{subfigure}
\usepackage{colortbl}
\usepackage{ulem}
\usepackage{longtable}

\newcommand\beq{\begin{equation}}
\newcommand\eeq{\end{equation}}
\newcommand\beqn{\begin{eqnarray}}
\newcommand\eeqn{\end{eqnarray}}

\begin{document}

\ensubject{Theoretical Physics}

\ArticleType{Article}
\Year{2018}
\Month{November}
\Vol{61}
\No{11}
\DOI{10.1007/s11433-018-9246-2}
\ArtNo{000000}
\ReceiveDate{April 8, 2018}
\AcceptDate{May 14, 2018}

\title{Localization of gravitino field on $f(R)$-thick branes}{Localization of gravitino field on $f(R)$-thick branes}

\author[1]{XiangNan Zhou}{}%
\author[2]{YunZhi Du}{}%
\author[3]{Hao Yu}{}%
\author[3]{YuXiao Liu}{{liuyx@lzu.edu.cn}}

\AuthorMark{X. N. Zhou}

\AuthorCitation{X. N. Zhou, Y. Z. Du, H. Yu, and Y. X. Liu}

\address[1]{College of Physics and Information Engineering,
  Shanxi Normal University, Linfen 041004, China;}
\address[2]{Institute of Theoretical Physics, Datong University, Datong 037009, China;}
\address[3]{Research Center of Gravitation and Institute of Theoretical Physics, Lanzhou University,
           Lanzhou 730000, China}


\abstract{In this paper, we consider the localization of a five-dimensional gravitino field on $f(R)$-thick branes. We obtain the coupled chiral equations of the Kaluza-Klein (KK) modes of gravitinos with the gauge condition $\Psi_z=0$. The chiral equations of a gravitino's KK modes are found to be almost identical to those of the Dirac fermion. However, their chiralities are exactly opposite. The chiral KK modes of gravitinos could be localized in some types of $f(R)$-thick branes on introducing a coupling term. We investigate the localization of a gravitino on three types of $f(R)$-thick branes through a Yukawa-like coupling term with background scalar fields. It has been shown that all the KK modes of gravitinos cannot be localized in the pure geometric $f(R)$-thick branes by adding a five-dimensional gravitino mass term. However, for the $f(R)$-thick branes generated by one or two background scalar fields, only the left- or right-handed zero mode could be localized in the branes, and the massive KK resonant modes are the same for both left- and right-handed gravitinos despite their opposing chiralities. All these results are consistent with those of the five-dimensional Dirac fermion except their chiralities, which may be an important sign to distinguish the gravitino field and the Dirac fermion field.}



\maketitle


\begin{multicols}{2}

\section{Introduction}
The extra-dimensional theory has attracted increasing attention even though the visible world is a four-dimension spacetime \cite{Akama:1982jy,Rubakov:1983,ADD,Antoniadis:1998ig,Randall:1999ee,Randall:1999vf,ArkaniHamed:2000eg,Kim:2000mc,Nussinov:2001rb,Pankov:2005ar,Abulencia:2006kk,Dey:2009xu,Neupane:2010ey}. Some classical physical problems including the gauge hierarchy problem (the huge difference between the Planck scale and the weak scale) \cite{Antoniadis:1998ig,Randall:1999ee,Das:2007qn,Yang:2012dd} and the cosmological problem \cite{Dvali:2000rv,Starkman:2001xu,Kim:2000mc,Dey:2009xu,Neupane:2010ey} could be solved by utilizing extra dimensions. In the 1920s, the Kaluza-Klein (KK) theory was proposed to unify Einstein's gravity and electromagnetism by introducing a compact extra spatial dimension with Planck size \cite{Kaluza,Klein}. Several decades later, Akama, Rubakov, and Shaposhnikov proposed the idea of a domain-wall braneworld with an extra infinite dimension in a five-dimensional flat spacetime \cite{Akama:1982jy,Rubakov:1983}. In 1998, Antoniadis and Arkani-Hamed et al. introduced a famous model with a large extra dimension in an attempt to solve the hierarchy problem \cite{ADD,Antoniadis:1998ig}. One year later, Randall and Sundrum (RS) suggested that this extra dimension with a warped geometry could be finite or infinite, corresponding to the RSI \cite{Randall:1999ee} or RSII \cite{Randall:1999vf} thin braneworld model. In both braneworld scenarios, our visible four-dimensional world is a brane {without} thickness along the extra dimension, and the matter fields of the standard model (SM) are confined to the brane, {whereas} only gravity propagates in the five-dimensional {bulk} spacetime. {Subsequently}, more realistic thick branes generated dynamically by matter fields or pure gravity were introduced \cite{Gregory:2000jc,Kaloper:2000jb,Melfo:2002wd,Bazeia:2004dh,Cardoso:2006nh,Liu:2011wi,Liu:2012rc,Liu:2012mia,Bazeia:2015owa,Yu:2015wma}. In these models, a nonvanishing distribution of energy density along the extra dimension existed.

{The braneworld {scenario} with} {warped} infinite extra dimensions {requires} a natural physical mechanism to trap the matter fields on the branes so as to not conflict with current experiments. Thus, investigating the localization of the matter fields on various types of branes is essential \cite{Chang:1999nh,Shiromizu:1999wj,Kehagias:2000au,Ringeval:2001cq,Maity:2003im,Chatterjee:2005cr,Melfo:2006hh,Liu:2007gk,Davies:2007xr,Liu:2009ve,Guerrero:2009ac,Fu:2012sa,Xie:2013rka,Liu:2013kxz,Zhao:2014gka,Guo:2014nja,Du:2015pjw}. To rebuild the SM on the branes, the zero modes of these matter fields (the four-dimensional massless particles) should be localized in the branes. Moreover, {the localization of massive KK modes is crucial to provide a method to explore extra dimensions.} For example, we may observe some physical effects when these KK particles interact with the SM particles in the Large Hadron Collider (LHC) \cite{Hung:2003cj,Guo:2011qt,Sahin:2014dua,Bauer:2016lbe}. In some braneworld models, there are no bounded massive KK modes for some matter fields, {whereas} there may be some resonant KK modes {quasilocalized} on the branes. These massive resonant KK modes may stay on the branes for a {long} time and interact with other particles, providing us with opportunities to find the massive resonant KK modes and prove the existence of extra dimensions \cite{Liu:2009ve,Xie:2013rka,Guo:2014nja,Almeida:2009jc,Liu:2009mg,Landim:2011ki,
Du:2013bx,Zhang:2016ksq}.

The gravitino is the gauge fermion supersymmetric partner of the graviton in the theory of supersymmetry.
It has been suggested as a candidate for dark matter in cosmology \cite{Chun:1993vz,Moroi:1995fs,Steffen:2006hw,Feng:2010ij,Savvidy:2012qa}.
It is a fermion of spin 3/2 and obeys the Rarita-Schwinger equation. The mass of {a} light gravitino is generally considered around 1 eV \cite{Chun:1993vz}, but there are still some challenges within the study of its mass \cite{Feng:2010ij}. Its mass has been widely investigated using the models of hot and cold dark matter \cite{Chun:1993vz,Steffen:2006hw}, and the possibility of finding light {gravitinos} at the LHC was discussed previously~\cite{Shirai:2009kn}. The behavior of a gravitino near a black hole has also gained attention \cite{Khlopov:2004tn,Yale:2008kx,Arnold:2013zva,Chen:2015jga}. Besides, a gravitino is a type of matter {field} beyond the SM and has many special properties that the SM matter fields do not possess.
Therefore, the localization of a five-dimensional gravitino field in a brane will be noteworthy, and provide us new perspectives for investigating the gravitino. Compared with the matter fields of the SM such as the scalar and fermion fields, {the works on the gravitino's fields are few and noncomprehensive} \cite{Liu:2007gk,Du:2015pjw,Bajc:1999mh,Oda:2000dd,Gherghetta:2000kr,Oda:2000wa,Hewett:2002uq,Lee:2007ib}. The zero mode of a {five-dimensional} free gravitino can be localized in a RS-like brane only when a bulk mass term is introduced \cite{Oda:2000dd}. In a $D$-dimensional spacetime with $D\geq5$, the zero mode of the gravitino with a coupling term can be localized in the brane, and its localization property is similar to that of the Dirac fermion \cite{Liu:2007gk,Oda:2000wa}. In addition, the behavior of the gravitino's KK modes with coupling terms was investigated previously~\cite{Hewett:2002uq}. Recently, the localization and mass spectrum of the gravitino's KK modes on two types of thin branes (the RS branes and the scalar-tensor branes) were investigated \cite{Du:2015pjw}. Note that most of these investigations focused on the RS-like thin branes.

In this paper, we focus on the localization of a five-dimensional gravitino field on the $f(R)$-thick branes. Although general relativity is a widely accepted theory, the problems of black holes, dark matter, and dark energy still attract people's attention \cite{Zhong:2016glr,Wei:2016,He:2017,Zhang:2017}. Furthermore, its nonrenormalization has motivated the investigation of modified gravity theories, particularly those involving high-order curvature terms \cite{Sotiriou:2008rp}. $f(R)$ gravity is a type of modified gravity whose Lagrangian is a function of the scalar curvature $R$. It always contains {high-order curvature invariants}, which could make the theory renormalizable \cite{Sotiriou:2008rp}. {Furthermore}, the $f(R)$ gravity could be used to explain the dark energy or dark matter and to answer astrophysical and cosmological paradoxes. Therefore, it has been studied widely in cosmology and the braneworld fields \cite{Liu:2011wi,Yu:2015wma,Sotiriou:2008rp,DeFelice:2010aj,Nojiri:2010wj,Nojiri:2000gv,Nojiri:2001ae,Giovannini:2001xg,Afonso:2007zz,Dzhunushaliev:2009dt,Liu:2011am,Bazeia:2015oqa,Zhong:2010ae,Zhong:2012nt,Zhong:2015pta}.
 Yu et al investigated various kinds of $f(R)$-branes and proposed general solutions~\cite{Yu:2015wma}. All of these solutions are also appropriate for general relativity braneworlds, i.e., $f(R)=R$.

In this study, we investigated the localization of a five-dimensional gravitino field on the $f(R)$-thick branes, whose solutions {have} been previously provided~\cite{Yu:2015wma}. The conclusions about the localization of the gravitino will have some universality because they are also appropriate for general-relativity braneworlds. We believe that we can obtain some interesting results for the structure of thick branes that are inapplicable for thin branes. Our work is organized as follows: In Section 2, we consider the localization of a five-dimensional, free, and massless gravitino field in a thick brane. We introduce the gauge condition {$\Psi_z=0$} and derive the Schr\"{o}dinger-like equations for {a gravitino's} KK modes. Then, we focus on the localization of a five-dimensional gravitino field with a coupling term on a thick brane in Section 3. Three types of $f(R)$-thick branes are considered, and the massive KK resonances are studied. Finally, the discussion and conclusion are presented in Section 4.

\section{Localization of a free gravitino's field on thick branes}
\label{sec2}

First, we consider the localization of a free massless gravitino's field on a thick brane in a five-dimensional spacetime. Usually, the five-dimensional line-element can be assumed as follows:
\begin{equation}
ds^2=g_{MN}dx^Mdx^N=\text{e}^{2A(y)}\hat g_{\mu\nu}(x)dx^\mu dx^\nu + dy^2.\label{Le}
\end{equation}
Here, $M$ and $N$ denote the curved five-dimensional spacetime indices, $\hat{g}_{\mu\nu}$ is the metric on the brane, and the warp factor $\text{e}^{2A(y)}$ is only the function of the extra dimension $y$. For convenience, the following coordinate transformation could be performed:
\begin{eqnarray}
dz=\text{e}^{-A(y)}dy,\label{ytoz}
\end{eqnarray}
which transforms the metric (\ref{Le}) as follows:
\begin{eqnarray}
ds^2=\text{e}^{2A(z)}(\hat g_{\mu\nu}dx^\mu dx^\nu + dz^2).\label{LE}
\end{eqnarray}

The action of a free, massless gravitino's field $\Psi$ in five-dimensional spacetime was previously given~\cite{Liu:2007gk,Oda:2000wa,Du:2015pjw}
\begin{eqnarray}
S_{\frac{3}{2}}=\int d^5 x \sqrt{-g}~\bar\Psi_M\Gamma^{[M}\Gamma^N\Gamma^{R]}D_N\Psi_R, \label{action}
\end{eqnarray}
and the corresponding equations of motion read as
\begin{eqnarray}
\Gamma^{[M}\Gamma^N\Gamma^{R]}D_N\Psi_R=0.\label{motionequation}
\end{eqnarray}
The Dirac gamma matrices $\Gamma^{M}$ in curved five-dimensional spacetime satisfy $\Gamma^M=e^M_{~~\bar M}\Gamma^{\bar M}$. $\Gamma^{\bar M}$ are the gamma matrices in flat five-dimensional spacetime, and $\{\Gamma^{\bar M},~\Gamma^{\bar N}\}=2\eta^{\bar M\bar N}$, where $\bar M$ and $\bar N$ represent the five-dimensional local Lorentz indices. The vielbein satisfies $g_{MN}=e_M^{~~\bar M}e_N^{~~\bar N}\eta_{\bar M\bar N}$, and for the metric (\ref{LE}), it is given by
\begin{eqnarray}
e_M^{~~\bar M}=\left(
\begin{array}{cc}
    \text{e}^{A}\hat e_\mu^{~~\bar\mu}&
    0\\
    0&
    \text{e}^{A}
\end{array}\right),~~~~
e^M_{~~\bar M}=\left(
\begin{array}{cc}
    \text{e}^{-A}\hat e^\mu_{~~\bar\mu}&
    0\\
    0&
    \text{e}^{-A}
\end{array}\right).
\end{eqnarray}
From the relations $e_{M\bar M}=g_{MN}e^{N}_{~~\bar M}$ and $e^{M\bar M}=g^{MN}e_{N}^{~~\bar M}$, we can get
\begin{eqnarray}
e_{M\bar M}=\left(
\begin{array}{cc}
\text{e}^{A}\hat e_{\mu\bar\mu}&0\\
0&\text{e}^{A}
\end{array}\right),~~~~~~
e^{M\bar M}=\left(
\begin{array}{cc}
\text{e}^{-A}\hat e^{\mu\bar\mu}&0\\
0&\text{e}^{-A}
\end{array}\right).
\end{eqnarray}
Thus, $\Gamma^M=\text{e}^{-A}(\hat e^\mu_{~~\bar\mu}\gamma^{\bar\mu},~\gamma^5)=\text{e}^{-A}(\gamma^{\mu},~\gamma^5)$, where $\gamma^{\mu}= \hat e^\mu_{~~\bar\mu}\gamma^{\bar\mu}$, $\gamma^{\bar\mu}$, and $\gamma^5$ are the flat gamma matrices in the four-dimensional Dirac representation. In this paper, we choose the following representation for the four-dimensional flat gamma matrices:
\begin{eqnarray}
\gamma^0=\left(
\begin{array}{cc}
0&-\text{i}\mathbb{I}\\
-\text{i}\mathbb{I}&0
\end{array}\right),~~
\gamma^i=\left(
\begin{array}{cc}
0&\text{i}\sigma^i\\
-\text{i}\sigma^i&0
\end{array}\right),~~
\gamma^5=\left(
\begin{array}{cc}
\mathbb{I}&0\\
0&-\mathbb{I}
\end{array}\right).\label{Gamma}
\end{eqnarray}
Here, $\mathbb{I}$ is a two-by-two unit matrix, and $\sigma^i$ are the Pauli matrices.
In this work, we only considered flat thick branes, i.e., $\hat{g}_{\mu\nu}=\eta_{\mu\nu}$. So we have $\hat{e}^{\mu}_{~\bar{\mu}}=\delta^{\mu}_{~\bar{\mu}}$ and $\gamma^{\mu}=\gamma^{\bar{\mu}}$. In addition, the covariant derivative of a gravitino's field is defined as
\begin{eqnarray}
D_N\Psi_R=\partial_N\Psi_R-\Gamma^M_{~~NR}\Psi_M+\omega_N\Psi_R,
\end{eqnarray}
where the spin connection $\omega_N$ is defined by
$\omega_N=\frac{1}{4}\omega_{N}^{~~\bar N\bar L}
          \Gamma_{\bar N}\Gamma_{\bar L}$
and $\omega_{N}^{~~\bar N\bar L}$ is given by
\begin{eqnarray}
\omega_{N}^{~~\bar N\bar L}
&=&\frac{1}{2}e^{M\bar N}(\partial_N e_M^{~~\bar L}-\partial_M e_N^{~~\bar L})
-\frac{1}{2}e^{M\bar L}(\partial_N e_M^{~~\bar N}-\partial_M e_N^{~~\bar N})\nonumber\\
&&-\frac{1}{2}e^{M\bar N}e^{P\bar L}(\partial_M e_{P\bar R}-\partial_P e_{M\bar R})e^{~\bar R}_N.
\end{eqnarray}
Thus, we obtain the nonvanishing components of $\omega_N$:
\begin{eqnarray}
\omega_\mu=\frac{1}{2}(\partial_z A)\gamma_\mu\gamma_5+\hat\omega_\mu.
\end{eqnarray}
Note that the four-dimensional spin connection $\hat\omega_\mu$ on a flat brane vanishes.
The nonvanishing components of $D_N\Psi_R$ are
\begin{eqnarray}
D_\mu\Psi_\nu
&=&\partial_\mu\Psi_\nu-\Gamma^M_{~~\mu\nu}\Psi_M+\omega_\mu\Psi_\nu\nonumber\\
&=&\hat D_\mu\Psi_\nu+(\partial_z A)\hat g_{\mu\nu}\Psi_z+\frac{1}{2}(\partial_z A)\gamma_\mu\gamma_5\Psi_\nu,\label{G1}\\
D_\mu\Psi_z
&=&\partial_\mu\Psi_z-\Gamma^M_{~~\mu z}\Psi_M+\omega_\mu\Psi_z\nonumber\\
&=&\partial_\mu\Psi_z-(\partial_z A)\Psi_\mu+\frac{1}{2}(\partial_z A)\gamma_\mu\gamma_5\Psi_z+\hat\omega_\mu\Psi_z,\label{G3}\\
D_z\Psi_\mu
&=&\partial_z\Psi_\mu-\Gamma^M_{~~z\mu}\Psi_M+\omega_z\Psi_\mu\nonumber\\
&=&\partial_z\Psi_\mu-(\partial_z A)\Psi_\mu,\label{G4}\\
D_z\Psi_z
&=&\partial_z\Psi_z-\Gamma^M_{~~zz}\Psi_M+\omega_z\Psi_z\nonumber\\
&=&\partial_z\Psi_z-(\partial_z A)\Psi_z.
\end{eqnarray}
Equation (\ref{motionequation}) includes five equations because $M$ extends over all five spacetime indices. There are two kinds of equations: $M=5$ and $M=\mu$. For the first case of $M=5$, the equation of motion reads as
\begin{eqnarray}
\Gamma^{[5}\Gamma^{N}\Gamma^{R]}D_N\Psi_R&=&
\Gamma^{[5}\Gamma^{\mu}\Gamma^{\nu]}D_\mu\Psi_\nu\nonumber\\
&=&\big([\Gamma^{\mu},~\Gamma^{\nu}]-g^{\mu\nu}\big)
    \Gamma^5\nonumber\\
    &&\times\Big(\hat D_\mu\Psi_\nu+(\partial_z A)\hat g_{\mu\nu}\Psi_z+\frac{1}{2}(\partial_z A)\gamma_\mu\gamma_5\Psi_\nu\Big) \nonumber\\
&=&0.\label{EOM1}
\end{eqnarray}
In this work, for convenience, we prefer to choose the gauge condition $\Psi_z=0$, with which we introduce the KK decomposition
\begin{eqnarray}
\Psi_\mu=\sum\psi^{(n)}_\mu(x)\xi_n(z),  \label{KKdecomposition}
\end{eqnarray}
where $\psi^{(n)}_\mu(x)$ is the four-dimensional gravitino field. Then, Eq. (\ref{EOM1}) is reduced to
\begin{eqnarray}
\big([\gamma^{\mu},~\gamma^{\nu}]-\hat g^{\mu\nu}\big)     \gamma^5~
     \Big(\hat D_\mu\psi_\nu^{(n)}
          +\frac{1}{2} (\partial_z A)\gamma_\mu\gamma_5\psi_\nu^{(n)}
     \Big)=0.  \label{EqsGravitino5}
\end{eqnarray}
For the four-dimensional massive gravitino field $\psi_\mu$, the following four equations should be satisfied \cite{Moroi:1995fs}
\begin{subequations}\label{Eqs4DGravitino}
\begin{eqnarray}
\gamma^{[\lambda}\gamma^{\mu}\gamma^{\nu]}\hat{D}_{\mu}\psi_{\nu}-m_{3/2}[\gamma^{\lambda},~\gamma^{\mu}]\psi_{\mu}&=&0, \\
\gamma^{\mu}\psi_{\mu}&=&0,\\
\hat{D}^{\mu}\psi_{\mu}&=&0,\\
(\gamma^{\mu}\hat{D}_{\mu}+m_{3/2})\psi_{\nu}&=&0.
\end{eqnarray}
\end{subequations}
Here, $m_{3/2}$ is the mass of a four-dimensional gravitino field $\psi_{\mu}$. Thus, the left-hand side of Eq. \eqref{EqsGravitino5} always vanishes for a four-dimensional gravitino field $\psi_{\mu}^{(n)}$ {satisfying} the above equation \eqref{Eqs4DGravitino}.
On the other hand, when we choose the gauge condition $\Psi_{z}=0$, the contribution of $\Gamma^{[5}\Gamma^{N}\Gamma^{R]}D_N\Psi_R$ in the five-dimensional gravitino action \eqref{action} vanishes; hence, Eq. (\ref{EOM1}) can be ignored. Then, we focus on the case of $M=\mu$, for which the equations of motion are
\begin{eqnarray}
&&\Gamma^{[\lambda}\Gamma^{N}\Gamma^{L]}D_N\Psi_L\nonumber\\
&=&
\Gamma^{[\lambda}\Gamma^{\mu}\Gamma^{\nu]}D_\mu\Psi_\nu+
\Gamma^{[\lambda}\Gamma^{\nu}\Gamma^{5]}D_\nu\Psi_z+
\Gamma^{[\lambda}\Gamma^{5}\Gamma^{\nu]}D_z\Psi_\nu\nonumber\\
&=&\text{e}^{-3A}\gamma^{[\lambda}\gamma^{\mu}\gamma^{\nu]}\hat{D}_{\mu}\Psi_{\nu}-\text{e}^{-3A}[\gamma^\lambda,~\gamma^\nu]\gamma_5
(\partial_z A+\partial_z)\Psi_{\nu}\nonumber\\
&=&0,
\label{EOM2}
\end{eqnarray}
where we used the gauge condition $\Psi_z=0$. When we introduce the decomposition \eqref{KKdecomposition} and consider the zero mode, which corresponds to the four-dimensional massless gravitino satisfying $\gamma^{[\lambda}\gamma^{\mu}\gamma^{\nu]}\hat{D}_{\mu}\psi^{(0)}_{\nu}=0$, we obtain the equation of motion for the extra-dimensional configuration $\xi_0(z)$:
\begin{eqnarray}
&& \gamma^{[\lambda}\gamma^{\mu}\gamma^{\nu]}\hat{D}_{\mu}\psi^{0}_{\nu}(x)\xi_0(z)
    -[\gamma^\lambda,~\gamma^\nu]\gamma_5\psi^{(0)}_\nu(x)
(\partial_z A+\partial_z)\xi_0(z)\nonumber\\
&=&-(\partial_z A+\partial_z)\xi_0(z)=0.
\end{eqnarray}
Obviously, the solution is
\begin{eqnarray}
\xi_0(z)=C\text{e}^{-A(z)},
\end{eqnarray}
where $C$ is a normalization constant. Substituting the zero mode $\xi_0(z)$ into the gravitino action (\ref{action}) yields
\begin{eqnarray}
S_{\frac{3}{2}}^{(0)} 
                 =\mathcal{I}_0
\int d^4 x~\sqrt{-\hat g}~\bar\psi^{(0)}_\lambda
 \gamma^{[\lambda}\gamma^\mu\gamma^{\nu]}\hat D_\mu\psi^{(0)}_\nu(x),
 \end{eqnarray}
where $\mathcal{I}_0  \equiv \int dz~\text{e}^{2A}\xi^2_0(z)=C^2\int dz =C^2\int e^{-A(y)}dy$. To localize the gravitino's spin 3/2 on a brane, the integral $\mathcal{I}_0$ must {be} finite. Therefore, only if we consider a RS-type brane model, the zero mode of a five-dimensional free massless gravitino can be localized in the brane for a finite extra dimension.

For the massive modes, we need to introduce the following chiral decomposition:
\begin{eqnarray}
\Psi_\mu(x,z)
&=&\sum_n\left(\psi^{(n)}_{L\mu}(x)\xi_{Ln}(z)+\psi^{(n)}_{R\mu}(x)\xi_{Rn}(z)\right) \nonumber \\
&=&\sum_n\bigg(\left[
\begin{array}{c}
0\\ \tilde{\psi}^{(n)}_{L\mu}\xi_{Ln}
\end{array}\right]+\left[
\begin{array}{c}
\tilde{\psi}^{(n)}_{R\mu}\xi_{Rn}\\0
\end{array}\right]
\bigg),\label{CD}
\end{eqnarray}
where $\tilde{\psi}^{(n)}_{L\mu}$ and $\tilde{\psi}^{(n)}_{R\mu}$ are both the two-component spinors. The effect of $P_{L,R}$ ($P_{L,R}=\frac{1}{2}[I \mp \gamma^5]$) on the gravitino field $\Psi_M$ is to differentiate the left- and right-handed parts, respectively, which are equivalent to the following equations:
\begin{eqnarray}
\gamma^5\psi^{(n)}_{L\mu}=-\psi^{(n)}_{L\mu}, ~~~~~~\gamma^5\psi^{(n)}_{R\mu}=\psi^{(n)}_{R\mu}. \label{partiyrelation}
\end{eqnarray}
Thus, substituting the chiral decomposition (\ref{CD}) into Eq. (\ref{EOM2}), we have
\begin{eqnarray}
&&\gamma^{[\lambda}\gamma^\mu\gamma^{\nu]}\hat D_\mu\psi^{(n)}_{L\nu}\xi_{Ln}
+\gamma^{[\lambda}\gamma^\mu\gamma^{\nu]}\hat D_\mu\psi^{(n)}_{R\nu}\xi_{Rn}
+[\gamma^{\lambda},~\gamma^{\nu}]
 (\partial_z A)\psi^{(n)}_{L\nu}\xi_{L n}\nonumber\\
&&-[\gamma^{\lambda},~\gamma^{\nu}](\partial_z A)\psi^{(n)}_{R\nu}\xi_{Rn}
+[\gamma^{\lambda},~\gamma^{\nu}]\psi^{(n)}_{L\nu}\partial_z\xi_{Ln}
-[\gamma^{\lambda},~\gamma^{\nu}]\psi^{(n)}_{R\nu}\partial_z\xi_{Rn}\nonumber\\
&=&0.
\end{eqnarray}
Because the product of three gamma matrices is an oblique diagonal and the product of two gamma matrices is diagonal, two equations can be obtained from above equation:
\begin{subequations}
\begin{eqnarray}
&&\gamma^{[\lambda}\gamma^\mu\gamma^{\nu]}\hat D_\mu\psi^{(n)}_{L\nu}\xi_{Ln}
-[\gamma^{\lambda},~\gamma^{\nu}](\partial_z A)\psi^{(n)}_{R\nu}\xi_{Rn}\nonumber\\
&&-[\gamma^{\lambda},~\gamma^{\nu}]\psi^{(n)}_{R\nu}\partial_z\xi_{Rn}=0,\\
&&\gamma^{[\lambda}\gamma^\mu\gamma^{\nu]}\hat D_\mu\psi^{(n)}_{R\nu}\xi_{Rn}
+[\gamma^{\lambda},~\gamma^{\nu}](\partial_z A)\psi^{(n)}_{L\nu}\xi_{L n}\nonumber\\
&&+[\gamma^{\lambda},~\gamma^{\nu}]\psi^{(n)}_{L\nu}\partial_z\xi_{Ln}=0.
\end{eqnarray}
\end{subequations}
Using the method of separation of variance and by defining a parameter $m_n$, we have
\begin{subequations}
\begin{eqnarray}
\frac{\gamma^{[\lambda}\gamma^\mu\gamma^{\nu]}\hat D_\mu\psi^{(n)}_{L\nu}}
     {[\gamma^{\lambda},~\gamma^{\alpha}]\psi^{(n)}_{R\alpha}}
=\frac{(\partial_z A)\xi_{Rn}+\partial_z\xi_{Rn}}{\xi_{Ln}}=m_n,\\
\frac{\gamma^{[\lambda}\gamma^\mu\gamma^{\nu]}\hat D_\mu\psi^{(n)}_{R\nu}}
     {[\gamma^{\lambda},~\gamma^{\alpha}]\psi^{(n)}_{L\alpha}}
=-\frac{(\partial_z A)\xi_{Ln}+\partial_z\xi_{Ln}}{\xi_{Rn}}=m_n,
\end{eqnarray}
\end{subequations}
i.e.,
\begin{eqnarray}
\gamma^{[\lambda}\gamma^\mu\gamma^{\nu]}\hat D_\mu\psi^{(n)}_{L\nu}
&=&m_n[\gamma^{\lambda},~\gamma^{\alpha}]\psi^{(n)}_{R\alpha},~~~~~~\nonumber\\
\gamma^{[\lambda}\gamma^\mu\gamma^{\nu]}\hat D_\mu\psi^{(n)}_{R\nu}
&=&m_n[\gamma^{\lambda},~\gamma^{\alpha}]\psi^{(n)}_{L\alpha},~~~~~~
\label{MLD}\\
(\partial_z+(\partial_z A))\xi_{Rn}&=&m_n\xi_{Ln},\nonumber\\
(\partial_z+(\partial_z A))\xi_{Ln}&=&-m_n\xi_{Rn}.~~
\label{MRD}
\end{eqnarray}
Equations (\ref{MLD}) are the ones that four-dimensional chiral gravitinos' fields satisfy, and Eqs. (\ref{MRD}) are the coupled ones that the KK modes $\xi_{Ln}$ and $\xi_{Rn}$ satisfy. Performing the field transformations  $\xi_{Rn}(z)=\chi^{R}_{n}(z)~\text{e}^{-A}$ and $\xi_{Ln}(z)=\chi^{L}_{n}(z)~\text{e}^{-A}$, we can obtain equations for the left- and right-handed KK modes of the gravitino
\begin{subequations}
\begin{eqnarray}
\partial_z^2\chi^{L}_n(z)=-m^2_n\chi^L_n(z),\label{chi1}\\
\partial_z^2\chi^{R}_n(z)=-m^2_n\chi^R_n(z).\label{chi2}
\end{eqnarray}
\end{subequations}
When the following normalizable conditions are introduced,
\begin{eqnarray}
\int\chi^{L}_{m}(z)\chi^{R}_n(z)dz=\delta_{RL}\delta_{mn}, \label{normalizable condition}
\end{eqnarray}
the effective action of the four-dimensional massless and massive gravitinos can be obtained
\begin{eqnarray}
S^{m}_{\frac{3}{2}}&=&\sum_{n}\int d^4x
       \bigg[\bar{\psi}^{(n)}_{L\lambda}(x)\gamma^{[\lambda}\gamma^{\mu}\gamma^{\nu]}\partial_{\mu}\psi^{(n)}_{L\nu}(x)
             -m_n\bar{\psi}^{(n)}_{L\lambda}(x)[\gamma^{\lambda},~\gamma^{\mu}]\psi^{(n)}_{R\mu}(x)\nonumber\\
          &&~~~~+\bar{\psi}^{(n)}_{R\lambda}(x)\gamma^{[\lambda}\gamma^{\mu}\gamma^{\nu]}\partial_{\mu}\psi^{(n)}_{R\nu}(x)
           -m_n\bar{\psi}^{(n)}_{R\lambda}(x)[\gamma^{\lambda},~\gamma^{\mu}]\psi^{(n)}_{L\mu}(x)
      \bigg]\nonumber\\
&=&\sum_{n}\int d^4x\left(\bar{\psi}^{(n)}_{\lambda}(x)\gamma^{[\lambda}\gamma^{\mu}\gamma^{\nu]}\partial_{\mu}\psi^{(n)}_{\nu}(x)
         -m_n\bar{\psi}^{(n)}_{\lambda}(x)[\gamma^{\lambda},~\gamma^{\mu}]\psi^{(n)}_{\mu}(x)\right).\nonumber\\\label{4Daction}
\end{eqnarray}
However, the solutions of Eqs. (\ref{chi1}) and (\ref{chi2}) are clearly mediocre. Thus, the four-dimensional massive gravitinos cannot be localized. This conclusion is the same as that for the Dirac fermion.

\section{Localization of gravitino field with the coupling term on thick branes}

As mentioned in the previous section, the massive KK modes of a five-dimensional free massless gravitino field cannot be localized in RS-type thick branes. Therefore, it is necessary to introduce a coupling term as that in the case of Dirac field. In the thin brane scenario \cite{Du:2015pjw}, one usually introduces an additional mass term that is associated with the warp factor of the thin brane. In the scenario of the thick brane generated by one or multiple background scalar fields, we can introduce a coupling term between the background scalar field and gravitino field. We consider the simplest coupling, i.e., a Yukawa-like coupling, for which the action of a five-dimensional gravitino field is
\begin{eqnarray}
S_{\frac{3}{2}}&=&\int d^5x\sqrt{-g}\nonumber\\
&&\Big(\bar{\Psi}_M\Gamma^{[M}\Gamma^{N}\Gamma^{R]}D_N\Psi_R-\eta F(\phi)\bar{\Psi}_M [\Gamma^M,~\Gamma^N]\Psi_N\Big).~~~~
\end{eqnarray}
Here, $F(\phi)$ is a function of the background scalar field $\phi$, and $\eta$ is the coupling constant. The equations of motion derived from the above action {are}
\begin{eqnarray}
\Gamma^{[M}\Gamma^N\Gamma^{R]}D_N\Psi_R-\eta F(\phi)[\Gamma^M,~\Gamma^N]\Psi_N=0.
\end{eqnarray}
By using the gauge condition $\Psi_z=0$ and introducing the chiral decomposition
\begin{eqnarray}
\Psi_\mu(x,z)=\sum_n\text{e}^{-A(z)}\left(\psi^{(n)}_{L\mu}(x)\chi^{L}_{n}(z)+\psi^{(n)}_{R\mu}(x)\chi^{R}_{n}(z)\right),
\end{eqnarray}
we can obtain the following first-order coupled equations
\begin{subequations}\label{FirstOrderCoupledEquations}
\begin{eqnarray}
(\partial_z-\eta\text{e}^{A}F(\phi))\chi^L_n(z)=&-&m_n\chi^R_n(z),\\
(\partial_z+\eta\text{e}^{A}F(\phi))\chi^R_n(z)=&&m_n\chi^L_n(z).
\end{eqnarray}
\end{subequations}
From the above equation \eqref{FirstOrderCoupledEquations}, the left- and right-handed KK modes of the gravitino field satisfy the following Schr\"{o}dinger-like equations:
\begin{subequations}\label{EoG}
\begin{eqnarray}
(-\partial^2_z+V^L(z))\chi^L_n(z)&=&m_n^2\chi^L_n(z),\label{EoL}\\
(-\partial^2_z+V^R(z))\chi^R_n(z)&=&m_n^2\chi^R_n(z),\label{EoR}
\end{eqnarray}
\end{subequations}
where the effective potentials are given by
\begin{subequations}\label{Vz}
\begin{eqnarray}
V^L(z)&=&(\eta\text{e}^{A}F(\phi))^2+\eta\partial_z(\text{e}^{A}F(\phi)),\\
V^R(z)&=&(\eta\text{e}^{A}F(\phi))^2-\eta\partial_z(\text{e}^{A}F(\phi)).
\end{eqnarray}
\end{subequations}
For a five-dimensional free gravitino, we have obtained the effective action (\ref{4Daction}) of the four-dimensional left- and right-handed gravitinos. It is interesting that the forms of these equations for the left- and right-handed KK gravitinos (\ref{EoL}) and (\ref{EoR}) are identical to those of the KK modes of a Dirac field; meanwhile, they only differ in terms of their chiralities. For a given background solution of a thick brane, if the function $F(\phi)$ and the coupling parameter $\eta$ are identical, the mass spectrum of the KK gravitinos will be identical to that of the Dirac field. Here, we should note the difference in chiralities, which will yield an interesting result.

Next, we review some kinds of $f(R)$-thick branes \cite{Zhong:2015pta,Yu:2015wma}, and then investigate the localization of the five-dimensional gravitino on these branes and present their KK mass spectra.

In the five-dimensional spacetime, the action of a general $f(R)$-thick brane model reads \cite{Yu:2015wma}
\begin{eqnarray}
S=\int d^{5}x\sqrt{-g}\left(\frac{1}{2\kappa^{2}_{5}}f(R)+L(\phi_{i},X_i)\right),
\end{eqnarray}
where $\kappa^{2}_{5}\equiv8\pi G^{5}$ is the five-dimensional gravitational constant and is set to one for convenience, $f(R)$ is a function of the scalar curvature $R$, and $L(\phi_{i},X_i)$ is the Lagrangian density of the background scalar fields $\phi_i$ with the kinetic terms $
 X_i=-\frac{1}{2}g^{MN}\partial_{M}\phi_{i}\partial_N\phi_{i}$.
It is predictable that the spectra of the KK modes of the gravitino field on these $f(R)$-thick branes will be almost the same as those of the Dirac field, except for their chiralities. These results could provide us some important references for the future experiments on extra dimensions and gravitinos.

\subsection{Localization of gravitino field on the pure geometric $f(R)$-thick {branes} without background scalar field}

First, we focus on the localization of the gravitino field on the pure geometric $f(R)$-thick branes. Zhong et al investigated pure geometric $f(R)$-thick branes, wherein the Lagrangian density of the background scalar fields $L(\phi_{i},X_i)$ vanishes~\cite{Zhong:2015pta}. For flat pure geometric $f(R)$-thick branes, the background metric was previously provided (\ref{Le}) with $\hat g_{\mu\nu}=\eta_{\mu\nu}$. The solution of the warp factor $A(y)$ is \cite{Zhong:2015pta}
\begin{eqnarray}
A(y)=-n\ln(\cosh(ky)), \label{Ay_fR1}
\end{eqnarray}
where $k$ is a positive real parameter related to the curvature of the five-dimensional spacetime, and $n$ is a positive integer. The solutions of the function $f(R)$ for $n=1$ and $n=20$ are respectively \cite{Zhong:2015pta}
\begin{eqnarray}
f(R)&=&\frac{1}{7}(6k^2+R)\cosh(a(w(R)))\nonumber\\
&&-\frac{2}{7}k^2\sqrt{480-\frac{36R}{k^2}-\frac{3R^2}{k^4}}\sinh(\alpha((w(R))),~(n=1)~~~~~~~~~~\\
f(R)&=&-\frac{377600}{7803}k^2+\frac{4196}{2601}R\nonumber\\
&&-\frac{83}{41616k^2}R^2+\frac{13}{39951360k^4}R^3,~(n=20)
\end{eqnarray}
where $\alpha(w)=2\sqrt{3}\arctan(\tanh(\frac{w}{2}))$ and $w(R)=\pm\text{arcsech}\left[\frac{\sqrt{20n^2+R/k^2}}{\sqrt{8n+20n^2}}\right]$.
For arbitrary $n$, the {function $f(R)$ has no unified expression}, and obtaining {an} analytical $y(z)$ from the following relation of $z(y)$ calculated from the solution \eqref{Ay_fR1} is difficult:
\begin{eqnarray}
 z(y) 
 =-\frac{\cosh^{n+1}(k y) \sinh(k y)
             ~{_2 F_1}\big(1/2, \frac{n+1}{2}, \frac{n+3}{2}, \cosh^2(k y)\big)}
          {(n+1) k \sqrt{-\sinh^2(k y)}}.~~~~~
 \end{eqnarray}
Because there is no background scalar {field} in the pure geometric brane model, we may try to take $\eta F$ as the five-dimensional mass $M$ of the gravitino field. Then, the effective potentials $V^L$ and $V^R$ can be expressed in terms of the extra dimension $y$.
 \begin{eqnarray}
 V^L(z(y))&=& \text{sech}^{2n}(ky) \big(M^2-nkM\tanh(ky)\big),\label{VLforpure}\\
 V^R(z(y))&=& \text{sech}^{2n}(ky) \big(M^2+nkM\tanh(ky)\big).  \label{VRforoure}
 \end{eqnarray}
It is easy to see that both potentials are asymmetric and that their asymptotic behaviors are
\begin{eqnarray}
V^L(0)&=&M^2,~V^L(\pm\infty)=\text{e}^{2A(\pm\infty)}(M^2\mp Mkn)=0,\\
V^R(0)&=&M^2,~V^R(\pm\infty)=\text{e}^{2A(\pm\infty)}(M^2\pm Mkn)=0,
\end{eqnarray}
which indicates that there is no bound massive KK {mode}. The solutions for the left- and right-handed zero modes of the gravitino field are $\chi_0^{L,R} \propto e^{\pm My}$. It is clear that both zero modes are not normalizable; hence, they cannot be localized in the pure geometric $f(R)$-thick branes.

\subsection{Localization of gravitino field on the $f(R)$ thick {branes} with $L=X-V(\phi)$}

Now let us consider the $f(R)$-thick branes generated by one background scalar field. For the Lagrangian density $L=X-V(\phi)=-\frac{1}{2}\partial^{M}\phi\partial_{M}\phi-V(\phi)$, the solution in this model with the Sine-Gordon potential is given by \cite{Yu:2015wma}
\begin{subequations}
\begin{eqnarray}
f({{\hat{R}}})
 \!&=&\!    \hat{R}
    +\!\alpha \bigg\{ \frac{24b^2\!+\!2{\hat{R}}\!+ \!2b{\hat{R}}}{2\!+\!5b}
     \Big[P_{K_{-}}^{{b}/{2}}(\Xi)
    \!-\!   {\beta}Q_{K_{-}}^{{b}/{2}}(\Xi)
     \Big]  \nonumber \\
    &-&\!\!
 4(b^2\!-\!2bK_{+}) \Xi\Big[P_{K_{+}}^{{b}/{2}}(\Xi)
     -\!\! \Xi P_{K_{-}}^{{b}/{2}}(\Xi) \nonumber \\
    &+&\!\!\beta \Xi\left( Q_{K_{-}}^{{b}/{2}}(\Xi)
    - Q_{K_{+}}^{{b}/{2}}(\Xi)\right)\Big] \bigg\}  \Theta^{b/2},  \label{fRBrane2a} \\
 V(\phi) \!&=&\! \frac{3b k^2}{8}\left[(1-4b)+(1+4b)\cos \Big(\sqrt{\frac{8}{3b}}\phi\Big)\right],~~~~\label{fRBrane2b}\\
\phi(y) \!&=&\! \sqrt{6b} \arctan \Big[\tanh\Big(\frac{ky}{2}\Big)\Big],\label{fRBrane2c}\\
A(y)\!&=&\! -b\ln\Big[\cosh (ky)\Big],\label{fRBrane2d}
\end{eqnarray}\label{fRBrane2}
\end{subequations}
where $b$ and $k$ are positive parameters related to the thickness of the brane; $\alpha$ is an arbitrary constant; $\hat{R}\equiv R/k^2$, $K_{\pm}\equiv\frac12\sqrt{(b-14) b+1}\pm1/2$; $\Xi=\sqrt{1-\Theta^2}$; $\Theta\equiv\frac{\sqrt{20b^2+{\hat{R}}}}{2\sqrt{2b+5b^2}}$; and $P$ and $Q$ are the first and second kinds of Legendre functions, $\beta={P_{K_{+}}^{{b}/{2}}(0)}/{Q_{K_{+}}^{{b}/{2}}(0)}$. Note that the solution \eqref{fRBrane2b}-\eqref{fRBrane2d} is also appropriate for the case of $f(R)=R$. Thus, the following results are also appropriate for the case of the general-relativity thick brane. As discussed in the above subsection, it is very difficult to obtain analytical $y(z)$. Therefore, in the following section, we will solve the equations numerically. The effective potentials $V^L$ and $V^R$ in the physical coordinate $y$ become
\begin{subequations}\label{Vzy}
\begin{eqnarray}
V^L(z(y))&=&(\eta\text{e}^{A}F(\phi))^2+\eta\text{e}^{2A}\partial_y F(\phi)+\eta(\partial_yA) \text{e}^{2A} F(\phi),~~~~~~~~~\\
V^R(z(y))&=&V^L(z(y))|_{\eta\rightarrow-\eta}.
\end{eqnarray}\label{Vforphi}
\end{subequations}
Obviously, {for different forms of $F(\phi)$, the potentials $V^L$ and $V^R$ have different expressions}, which determine the mass spectra of the KK modes. In this work, we consider one kind of Yukawa coupling, i.e., $F(\phi)=\phi^\alpha$ with positive integer $\alpha$. For a kink configuration of the scalar $\phi$, because $V^L$ and $V^R$ should be symmetrical with respect to the extra dimension $y$, $\alpha$ should be odd. Next we consider two cases: the simplest case $F(\phi)=\phi$ and the case for $\alpha>1$.


\begin{figure*}[!htb]
\begin{center}
\subfigure[$V^L$, $b=1$]{\label{fig:VLRCaseII1a}
\includegraphics[width=0.4\textwidth]{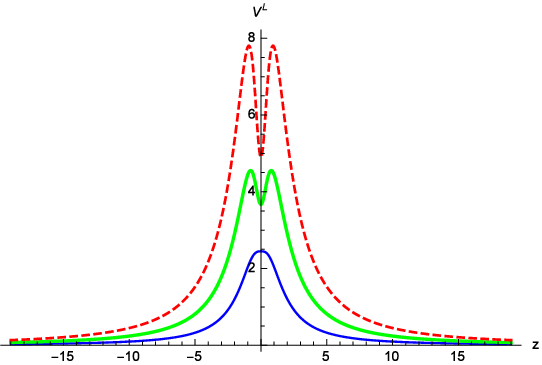}}
\subfigure[$V^R$, $b=1$]{\label{fig:VLRCaseII1b}
\includegraphics[width=0.4\textwidth]{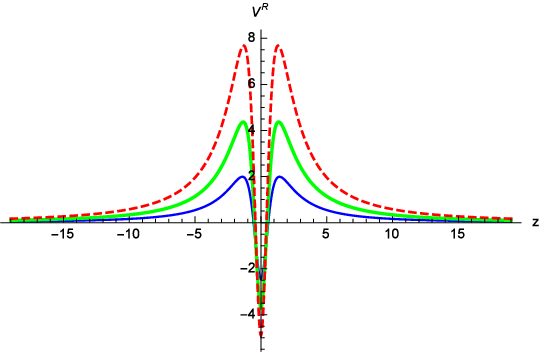}}
\subfigure[$V^L$, $b=3$]{\label{fig:VLRCaseII1c}
\includegraphics[width=0.4\textwidth]{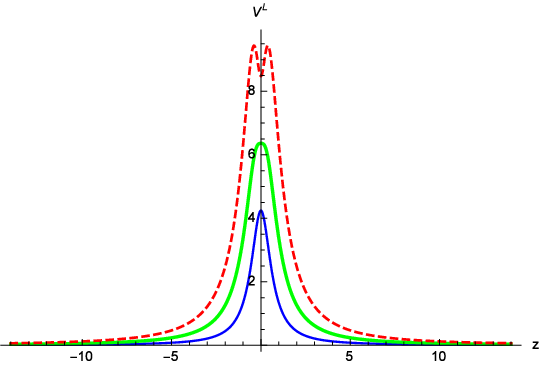}}
\subfigure[$V^R$, $b=3$]{\label{fig:VLRCaseII1d}
\includegraphics[width=0.4\textwidth]{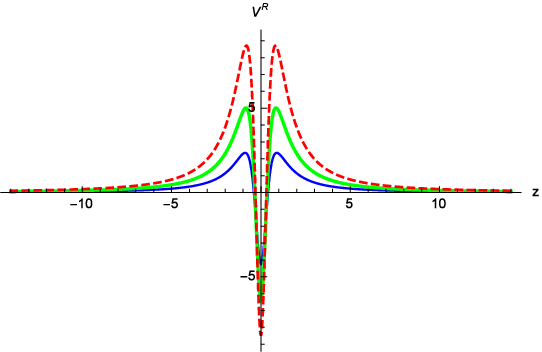}}
\end{center} \vskip -5mm
\caption{Potentials $V^{L}(z)$ and $V^{R}(z)$ for the left- and right-handed gravitinos on the $f(R)$-thick branes with $F(\phi)=\phi$. Here, $k=1$, and the coupling constant $\eta$ is set to $2.0$ (blue thin trace), $3.0$ (green thick trace), and $4.0$ (red dashed trace).}
 \label{fig:VLRCaseII1}
\end{figure*}

\subsubsection{Case I: $F(\phi)=\phi$}
For the case of $F(\phi)=\phi$, the effective potentials (\ref{Vforphi}) read
\begin{subequations}
\begin{eqnarray}
 V^L(y)&=&\frac{1}{2}\cosh(ky)^{-1-2b}
   \bigg [ 12b\eta^2\arctan\Big(\tanh(\frac{ky}{2})\Big)^2\cosh(ky)\nonumber\\
       &+&\eta\sqrt{6b}k(1-2b\arctan\Big(\tanh(\frac{ky}{2})\Big)\sinh(ky))  \bigg] ,
       \label{Vforphi1L} \\
 V^R(y)&=&V^L(y)|_{\eta\rightarrow-\eta},\label{Vforphi1R}
\end{eqnarray}\label{Vforphi1}
\end{subequations}
which are symmetrical. The values of the potentials at the origin and at infinity are given by
\begin{eqnarray}
&&V^R(0)=-\eta k\sqrt{\frac{3b}{2}}=-V^L(0),\\
&&V^R(\pm\infty)=0=V^L(\pm\infty).
\end{eqnarray}
It is clear that both potentials exhibit the same asymptotic behaviors as $y\rightarrow\pm\infty$, whereas their values at $y=0$ are opposite. Thus, only the left- or right-handed gravitino zero mode (four-dimensional massless left- or right-handed gravitino) could be localized in the $f(R)$-thick brane. The shapes of the potentials (\ref{Vforphi1}) are shown in Fig.~\ref{fig:VLRCaseII1}; it can be seen that for any positive $b$, $k$, and $\eta$, $V^R(z(y))$ is a volcano-type potential and that a localized zero mode and a continuous gapless spectrum of massive KK modes may exist. Furthermore, the depth of the potential $V^R$ increases with values of the parameters $\eta$, $b$, and $k$. By solving Eq.~(\ref{EoR}) with the potential (\ref{Vforphi1R}), the zero mode of the right-handed gravitino {becomes}
\begin{eqnarray}
\chi^{R}_{0}(z)&\propto&\exp\left(-\eta\int_{0}^{z}e^{A(\bar{z})}
       F(\phi)d\bar{z}\right) \nonumber \\
 &=& \exp\left( -\eta\int_{0}^{y}\phi(\bar{y}) d\bar{y}\right)\nonumber\\
 &=& \exp\left( -\eta\int_{0}^{y}\sqrt{6b} \arctan \left(\tanh\left(\frac{k\bar{y}}{2}\right)\right) d\bar{y}\right),
       \label{fL0CaseI}
\end{eqnarray}
and its normalization condition
\begin{eqnarray}
 \int_{-\infty}^{\infty}(\chi^R_{0}(z))^2 dz
 &=& \int_{-\infty}^{\infty}(\chi^R_{0}(y))^2 e^{-A(y)} dy \nonumber \\
 &\propto& \int_{-\infty}^{\infty}  \exp\left( -A(y)-2\eta\int_{0}^{y} \phi(\bar{y}) d\bar{y}\right)dy \nonumber \\
 &=& \int_{-\infty}^{\infty}  \exp\Big( b
    \text{ln}(\cosh(ky))\nonumber \\
   &&-2\eta\int_{0}^{y}\sqrt{6b} \arctan \left(\tanh\left(\frac{k\bar{y}}{2}\right)\right) d\bar{y}
   \Big)dy
 <\infty~~~~~~~
 \end{eqnarray}
 is equivalent to
\begin{eqnarray}
 \int_{0}^{\infty}  \exp
   \left( kby    - \frac{\pi\eta}{2}\sqrt{6b}y \right)dy
 <\infty
       \label{NormalizationConditionfR0CaseI1}
\end{eqnarray}
because $-A(y)\rightarrow kby$ and $\arctan(\tanh(\frac{ky}{2}))={\pi}/{4}$ as $y\rightarrow\infty$. The above normalization condition (\ref{NormalizationConditionfR0CaseI1}) requires
\begin{eqnarray}
\eta>\eta_0\equiv\frac{k}{\pi}\sqrt{\frac{2b}{3}}.
       \label{NormalizationConditionfR0CaseI1Foreta}
\end{eqnarray}
Thus, if the coupling constant is strong enough ($\eta>\eta_0$), the right-handed zero mode can be localized in the brane. It is not difficult to check whether the left-handed zero mode canbe localized in the brane under the condition \eqref{NormalizationConditionfR0CaseI1Foreta}.

On the other hand, the potential $V^L(z(y))$ for positive $\eta$ is always positive and vanishes with increasing distance from the brane. This type of potential cannot trap any bound state; hence, there is no left-handed gravitino zero mode. The structure of the potential $V^L$ is determined by the parameters $k$, $b$, and $\eta$. For given $k$ and $b$, the potential $V^L$ has a barrier for a small $\eta$. When $\eta$ increases, a quasipotential well appears, and the depth of the well increases with the value of $\eta$. However, for given $\eta$ and $k$ (or $b$), the height of the potential $V^L$ increases with $b$ (or $k$), and the quasipotential well changes into a barrier with the growth of $b$ (or $k$). The behavior of $V^L$ around the point $y=0$ is similar to that of the function $y^4$, and there will be three extreme points if a quasipotential well exists around the point $y=0$. Using the third-order Taylor series expansion of $\partial_y V^L$ near the point $y=0$, we obtain
\begin{eqnarray}
 \partial_y V^L &=&
     \frac{1}{2} k^2 \eta \big[6b \eta-\sqrt{6b}k(1 + 4b)\big] y  \nonumber \\
    &+&\frac{1}{12} k^4 \eta \big[\sqrt{6b}k(1 + 2b) (5 + 18b) -24b\eta (1 + 3b) \big] y^3 + \mathcal{O}(z^5).~~~~~~~~~
\end{eqnarray}
For $k=1$ and $b>\frac{1}{2\sqrt{3}}$, the above function has three roots, and a quasipotential well appears when $\eta>\frac{1}{6}\sqrt{\frac{6+48b+96b^2}{b}}$ (it equals 2.04124 when $b=1$).

For the case wherein a quasipotential well appears for $V^L$, we can find the resonance states of the gravitino, which are massive four-dimensional gravitinos with finite lifetimes on the brane. To investigate the gravitino's resonant modes, we define the relative probability as follows~\cite{Liu:2009ve}:
\begin{eqnarray}
 P_{L,R}(m^{2})=\frac{\int^{z_{b}}_{-z_{b}}|\chi^{L,R}(z)|^{2}dz}{\int^{z_{max}}_{-z_{max}}|\chi^{L,R}(z)|^{2}dz},
\end{eqnarray}
where $2z_b$ is approximately the width of the brane, and $z_{max}=10z_b$. The left- and right-handed wavefunctions $\chi^{L,R}(z)$ are the solutions of Eqs.~(\ref{EoG}). The above definition can explain that $|\chi^{L,R}(z)|^{2}$ is the probability density ~\cite{Liu:2009ve,Almeida:2009jc}. A resonant mode with mass $m_n$ exists if the relative probability $P(m^2)$ has a peak around $m=m_n$. These peaks should have full width at half maximum, and the number of these peaks is the same as that of the resonant modes. To obtain the solutions of Eqs.~(\ref{EoG}), we always need two additional types of initial conditions
\begin{subequations}
\begin{eqnarray}
\label{even}
\chi^{L,R}_{\rm{even}}(0)\!\!&=&\!\!1, ~~~\partial_{z}\chi^{L,R}_{\rm{even}}(0)=0;\\
\label{odd}
\chi^{L,R}_{\rm{odd}}(0)\!\!&=&\!\!0, ~~~~\partial_{z}\chi^{L,R}_{\rm{odd}}(0)=1,
\end{eqnarray}\label{EvenOddConditions}
\end{subequations}
where $\chi^{L,R}_{\rm{even}}$ and $\chi^{L,R}_{\rm{odd}}$ correspond to the even- and odd-parity modes of $\chi^{L,R}(z)$, respectively.

\begin{figure*}[!htb]
\begin{center}
 \includegraphics[width=7cm,height=5cm]{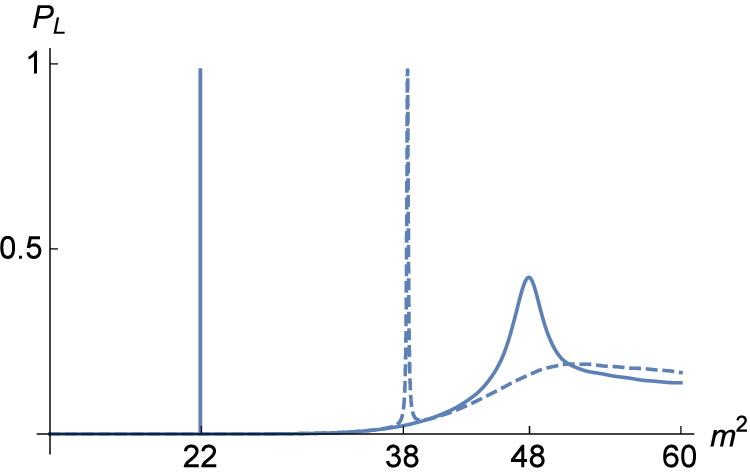}
 \includegraphics[width=7cm,height=5cm]{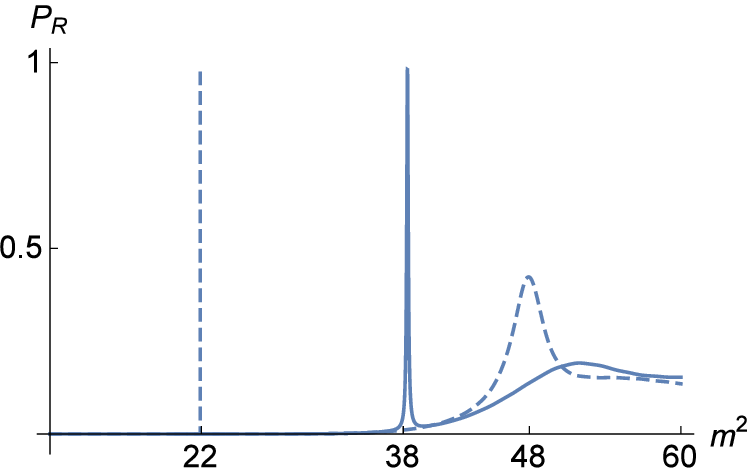}
\end{center} \vskip -5mm
\caption{The probabilities $P_{L,R}$ (as a function of $m^{2}$)
 for finding massive resonant KK modes of the left- and right-handed
 gravitinos with mass $m^{2}$ on the thick brane
 for the coupling $F(\phi)=\phi$.
 Solid lines and dashed lines are plotted for the even-parity
 and odd-parity massive gravitinos, respectively. The parameters are set to $b=1$, $k=1$, $\eta=10$, and $z_{max} = 20$.}
 \label{Pforphi}
\end{figure*}

\begin{figure*}[!htb]
\begin{center}
 \subfigure[$m^{2}=21.9171$]{\label{fig_fL_Eigenvalue_1a}
  \includegraphics[width=4.5cm,height=3.5cm]{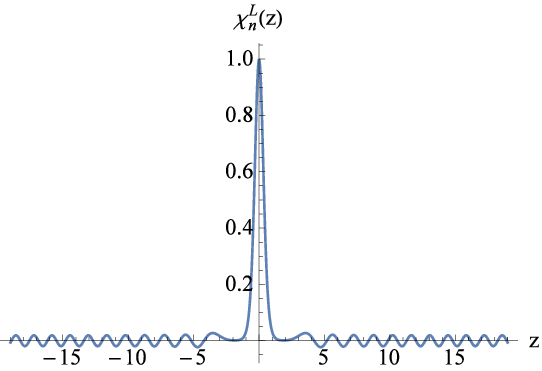}}
 \subfigure[$m^{2}=38.3348$]  {\label{fig_fL_Eigenvalue_1b}
  \includegraphics[width=4.5cm,height=3.5cm]{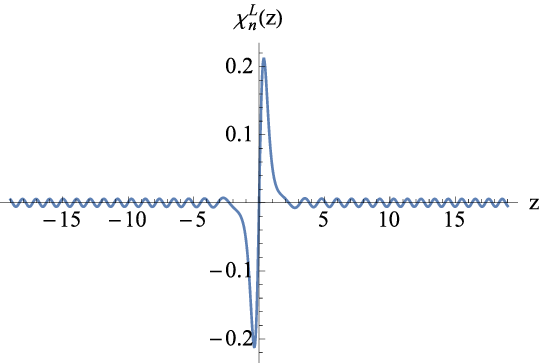}}
 \subfigure[$m^{2}=47.9467$]   {\label{fig_fL_Eigenvalue_1c}
  \includegraphics[width=4.5cm,height=3.5cm]{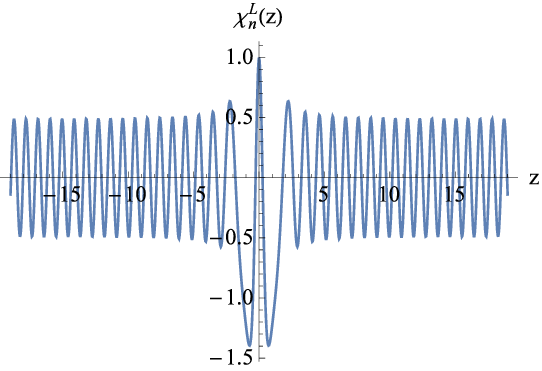}}
 \subfigure[$m^{2}=21.9169$]    {\label{fig_fL_Eigenvalue_1d}
  \includegraphics[width=4.5cm,height=3.5cm]{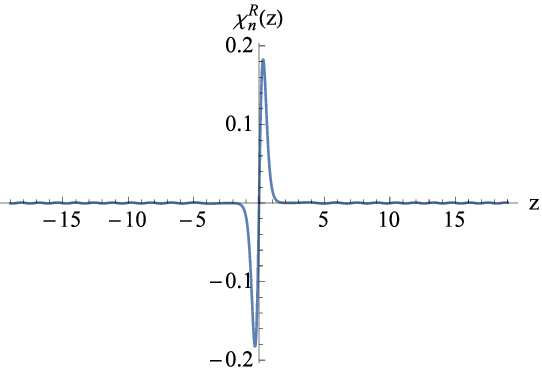}}
 \subfigure[$m^{2}=38.3311$]     {\label{fig_fL_Eigenvalue_1e}
  \includegraphics[width=4.5cm,height=3.5cm]{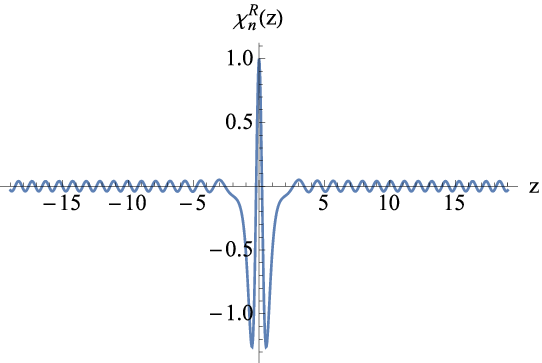}}
 \subfigure[$m^{2}=47.9328$]     {\label{fig_fL_Eigenvalue_1e}
  \includegraphics[width=4.5cm,height=3.5cm]{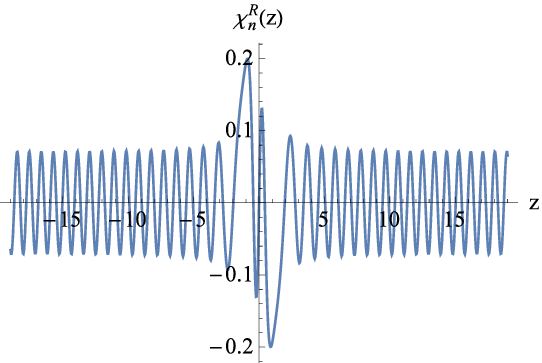}}
\end{center} \vskip -5mm
\caption{The shapes of the massive KK resonant modes of the left-handed (upper) and right-handed (lower) gravitinos for coupling $F(\phi)=\phi$ with different $m^2$.
Here, the parameters are set to $k=1$, $b=1$, $\eta=10$, and $z_{max} = 20$.}
 \label{fig__Eigenvalue1}
\end{figure*}

Our results are shown in Figs.~\ref{Pforphi}, \ref{fig__Eigenvalue1}, and Table~\ref{Tableonephi}. Obviously, the mass spectra of left- and right-handed gravitinos' resonant modes are almost the same, whereas their parities are opposite. The first resonant mode of the left-handed gravitino is even and its shape around $z=0$ looks like a ground state. Conversely, the first resonant mode of the right-handed gravitino is odd and it appears to be the first excited state. These results are reasonable because the effective potentials $V^L$ and $V^R$ are supersymmetric partners, which give the same spectra for resonant modes.
In fact, fermion resonances on branes have similar properties because Eqs.~(\ref{EoG}) of the KK modes of a gravitino are almost the same as those of a fermion. However, there is a difference between them, which is elaborated as follows: for a five-dimensional Dirac fermion field with a coupling term, if we use the representation of the gamma matrices (\ref{Gamma}) and parity relation (\ref{partiyrelation}), the equations of motion of the left- and right-handed fermion KK modes $f^{L,R}$ are given by
\begin{subequations}\label{Scheq}
\begin{eqnarray}
(-\partial_{z}^{2}+V^{L}(z))f^{L}&=&m^{2}f^{L},\label{ScheqLeft}
    \\
 (-\partial_{z}^{2}+V^{R}(z))f^{R}&=&m^{2}f^{R},
       \label{ScheqRight}
\end{eqnarray}
\end{subequations}
with the effective potentials
\begin{subequations}\label{Vz}
\begin{eqnarray}
V^{L}(z)=\eta^{2} e^{2A}F^{2}(\phi)-\eta
e^{A}\partial_{z}F(\phi)-\eta
e^{A}(\partial_{z}A)F(\phi)\,,  \label{VzL}  \\
V^{R}(z)=\eta^{2} e^{2A}F^{2}(\phi)+\eta
e^{A}\partial_{z}F(\phi)+\eta
e^{A}(\partial_{z}A)F(\phi)\,.   \label{VzR}
\end{eqnarray}
\end{subequations}
Obviously, the Schr\"{o}dinger-like equation of the left-handed gravitino's KK modes (\ref{EoL}) is the same as that of the right-handed fermion's KK modes (\ref{ScheqRight}), and the Schr\"{o}dinger-like equation for the right-handed gravitino's KK modes (\ref{EoR}) is the same as that of the left-handed fermion's KK modes (\ref{ScheqLeft}). Therefore, for a five-dimensional Dirac fermion, only the zero mode of the left-handed fermion can be localized in the $f(R)$-thick brane with the coupling $F(\phi)=\phi$, and the first resonant mode of the right-handed fermion is even. This difference between the fermion's and the gravitino's KK modes is generated from the differences in their field equations. For a five-dimensional Dirac fermion field with the Yukawa coupling, the field equation reads
\begin{equation}
\left[\gamma^{\mu}\partial_{\mu}+\gamma^5(\partial_z+2\partial_zA)-\eta \text{e}^{A}F(\phi)\right]\Psi=0.
\end{equation}
Note that the sign in front of $\gamma^5$ is positive. For bulk gravitino, Eq. (\ref{EOM2}) proves that the sign in front of $\gamma^5$ is negative, which leads to the swapping of the above results. This difference is noteworthy, and it could mark the distinction between the Dirac fermion and gravitino fields.

In addition, the number of the resonant modes for the gravitino field increases with the coupling constant $\eta$ but decreases with the parameter $b$. The relative probability $P$ decreases when the mass of the resonant mode approaches the maximum value of the potentials. Furthermore, the resonant modes approach each other as $m^2$ approaches the maximum value of the potentials. These results are consistent with those of the Dirac fermion.

\begin{table*}[!h]
\begin{center}
\begin{tabular}{||c|c|c|c|c|c|c||}
 \hline
 ~~~$b$~~~ & ~~~$\eta$~~~  & ~~~$\mathcal{C}$~~~ & ~~~$\mathcal {P}$~~~  & ~~~~~~~$m^2$ ~~~~~~~& ~~~~~~~$m$ ~~~~~~~&  ~~~~~~~$P$~~~~~~~  \\
 \hline\hline

                           &    &                   & even& 21.9171      & 4.68157     & 0.988172                                         \\ \cline{4-7}
                           &    & $\mathcal{L}$     & odd & 38.3348      & 6.19151     & 0.986425                                         \\ \cline{4-7}
                           &10  &                   & even& 47.9467      & 6.92436     & 0.423099                                         \\ \cline{3-7}
                           &    &                   & odd & 21.9169      & 4.68155     & 0.999673                                         \\ \cline{4-7}
                           &    &  $\mathcal{R}$    & even& 38.3311      & 6.19121     & 0.981257                                         \\ \cline{4-7}
                           &    &                   &odd  & 47.9328      & 6.92335     & 0.423111                                         \\ \cline{2-7} \cline{2-7}

                           &    &                   &even & 34.2006      & 5.84813     & 0.999212                                         \\ \cline{4-7}
                           &    &                   &odd  & 63.1603      & 7.94735     & 0.999998                                         \\ \cline{4-7}
\raisebox{2.3ex}[0pt]{1}   &    &  $\mathcal{L}$    &even & 86.4700      & 9.29892     & 0.993796                                         \\ \cline{4-7}
                           &    &                   &odd  & 102.7540     & 10.13680    & 0.632409                                         \\ \cline{4-7}
                           & 15 &                   &even & 112.2329     & 10.59400    & 0.238222                                         \\ \cline{3-7}
                           &    &                   &odd  & 34.1964      & 5.84777     & 0.999959                                         \\ \cline{4-7}
                           &    &                   &even & 63.1822      & 7.94872     & 0.992303                                         \\ \cline{4-7}
                           &    &   $\mathcal{R}$   &odd  & 86.4490      & 9.29780     & 0.998145                                         \\ \cline{4-7}
                           &    &                   &even & 102.7740     & 10.13780    & 0.631364                                         \\ \cline{4-7}
                           &    &                   &odd  & 112.0613     & 10.58590    & 0.232899                                         \\ \hline \hline

                           &    &   $\mathcal{L}$   &even & 35.3730      & 5.94752     & 0.982136                                         \\ \cline{4-7}
                           & 10 &                   &odd  & 54.7429      & 7.39884     & 0.287870                                         \\ \cline{3-7}
                           &    &   $\mathcal{R}$   &odd  & 35.3712      & 5.94737     & 0.981886                                         \\ \cline{4-7}
                           &    &                   &even & 54.4129      & 7.37651     & 0.285428                                         \\ \cline{2-7}\cline{2-7}

                           &    &                   &even & 56.8585      & 7.54046     & 0.999979                                         \\ \cline{4-7}
                           &    & $\mathcal{L}$     &odd  & 98.8432      & 9.94199     & 0.922543                                         \\ \cline{4-7}
\raisebox{2.3ex}[0pt]{3}   & 15 &                   &even & 122.4728     & 11.06670    & 0.276112                                         \\ \cline{3-7}
                           &    &                   &odd  & 56.8515      & 7.53999     & 0.999923                                         \\ \cline{4-7}
                           &    & $\mathcal{R}$     &even & 98.9253      & 9.94612     & 0.902362                                         \\ \cline{4-7}
                           &    &                   &odd  & 122.6000     & 11.07250    & 0.277316                                         \\ \hline
\end{tabular}\\
\caption{The eigenvalue $m^2$, mass $m$, and the relative probability of the left- and right-handed gravitinos with odd- and even-parity solutions for the coupling $F(\phi)=\phi$. In all tables presented in this paper, $\mathcal {C}$ and $\mathcal {P}$ represent chirality and parity, respectively, and $\mathcal {L}$ and $\mathcal {R}$ represent left- and right-handed, respectively. The parameter $k$ is set to $k=1$.}
\label{Tableonephi}
\end{center}
\end{table*}

\subsubsection{Case II: $F(\phi)=\phi^\alpha$ with $\alpha>1$}

\begin{figure*}[!htb]
\begin{center}
\subfigure[$V^L$, $\alpha=3$]{\label{fig:VLRCaseII1a}
\includegraphics[width=7cm,height=4.5cm]{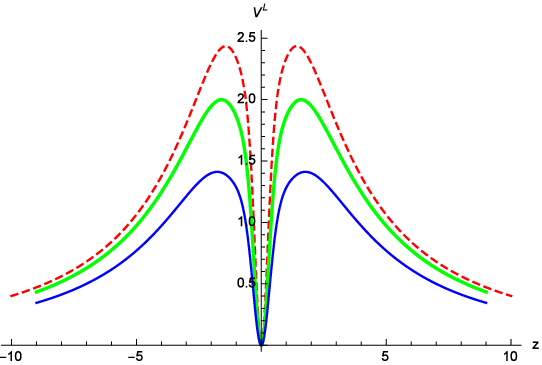}}
\subfigure[$V^R$, $\alpha=3$]{\label{fig:VLRCaseII1b}
\includegraphics[width=7cm,height=4.5cm]{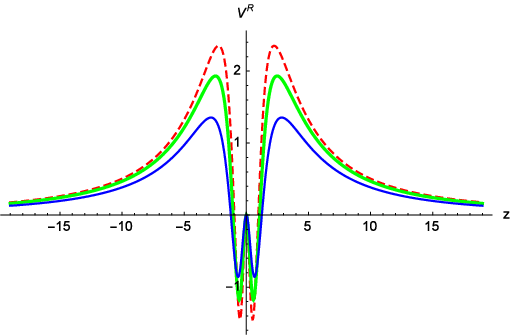}}
\subfigure[$V^L$, $\alpha=5$]{\label{fig:VLRCaseII1c}
\includegraphics[width=7cm,height=4.5cm]{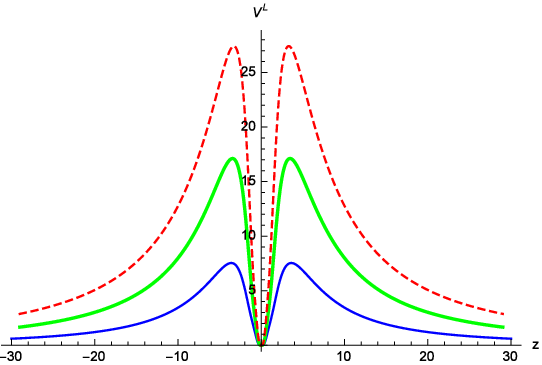}}
\subfigure[$V^R$, $\alpha=5$]{\label{fig:VLRCaseII1d}
\includegraphics[width=7cm,height=4.5cm]{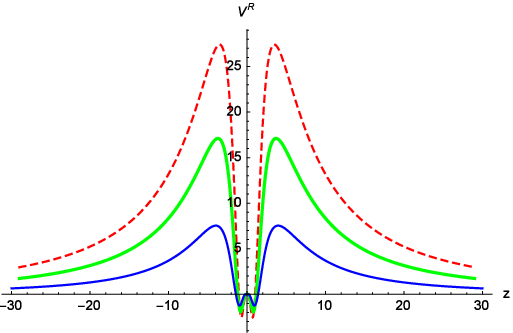}}
\end{center} \vskip -5mm
\caption{Potentials $V^{L}(z)$ and $V^{R}(z)$ for left- and right-handed gravitinos on the $f(R)$-thick branes with $F(\phi)=\phi^{\alpha}$. Here, $k=1$; $\eta=1$; and $b$ is set to $1.0$ (blue thin trace), $1.5$ (green thick trace), and $2.0$ (red dashed trace).}
 \label{fig:VLRCaseII2}
\end{figure*}

Next, we consider a natural generalization of the Yukawa coupling $F(\phi)=\phi^\alpha$ with $\alpha=3,~5,~7,~\cdots$. Note that $\phi^\alpha$ becomes a double kink for $\alpha\geq3$ because the scalar field $\phi$ is a kink. For this case, the effective potentials (\ref{Vforphi}) become
\begin{subequations}
\begin{eqnarray}
V^L(y)&=& \frac{1}{2} 3^{\frac{\alpha}{2}}  k\eta b^{\frac{\alpha}{2}-1}
        \arctan^{\alpha-1}\left(\tanh\left(ky/2\right)\right)
     \text{sech}^{2b+1}(k y)\nonumber\\
     &&\times\left[\alpha-2b\arctan\left(\tanh(ky/2)\right)\sinh(ky)\right] \nonumber\\
      &+& 6^\alpha\eta^2\left(\sqrt{b}\arctan\left(\tanh(ky/2)\right)\right)^{2\alpha}
      \text{sech}^{2b}(ky),\\
 V^R(y)&=&V^L(y)|_{\eta\rightarrow-\eta}.
\end{eqnarray}\label{Vforphi2}
\end{subequations}
Obviously, both the potentials are symmetrical and vanish at $y=0$ and $y\rightarrow\pm\infty$, and they are depicted in Fig.~\ref{fig:VLRCaseII2} for different values of $b$ and $\alpha$. There always exists a quasipotential well for the left-handed potential $V^L$ and a double-potential well for the right-handed one. These wells for both potentials increase in depth with {increases} in the parameters $b$, $\eta$, and $\alpha$, implying that there is an increasing number of resonances with increases in $b$, $\eta$, and $\alpha$. Because the coupling function $\phi^\alpha$ tends to a constant as $y\rightarrow\pm\infty$, the zero mode of the right-handed gravitino
\begin{eqnarray}
\chi^{R}_{0}\propto\exp\left(-\eta\int_{0}^{z}e^{A(z)}
       \phi^{\alpha}d z\right)
 =\exp\left( -\eta\int_{0}^{y}\phi^{\alpha}d y\right)
       \label{fL0CaseI}
\end{eqnarray}
is equivalent to $ \exp\left(-\eta(\frac{\pi}{4}\sqrt{6b})^\alpha|y| \right)$ because $\phi^\alpha=\pm(\frac{\pi}{4}\sqrt{6b})^\alpha$ as  $y\rightarrow\pm\infty$. The satisfaction of the normalization condition is easy to check for any positive coupling constant $\eta$. Thus, the right-handed zero mode can be localized in the brane for any positive coupling constant {$\eta$}; simultaneously, the {left-handed} zero mode cannot be localized.

For the massive modes, we consider the resonance states. As in the previous subsection, we solve the Schr\"{o}dinger equations (\ref{EoG}) numerically by using the two types of initial conditions (\ref{EvenOddConditions}). The mass spectrum of the resonances is presented in Table~\ref{Tableonephi3}.
It is clear that in this table, the masses of the resonant modes of the left- and right-handed gravitinos are still almost the same, whereas their parities are opposite. The number of the resonances increases with an increase in the parameters $b$, $\alpha$, and $\eta$. These resonances approach each other as $m^2$ increases, which is the same as the conclusion in the case of $\alpha=1$.

\begin{table*}[!h]
\begin{center}
\begin{tabular}{||c|c|c|c|c||}
 \hline
 ~~~$\alpha$~~~   & ~~~$\mathcal{C}$~~~ & ~~~$\mathcal {P}$~~~  & ~~~~~~~$m^2$ ~~~~~~~& ~~~~~~~$m$ ~~~~~~~ \\
 \hline\hline

                           & $\mathcal{L}$     & even & 1.07851      & 1.03851                                            \\ \cline{2-5}
 \raisebox{2.3ex}[0pt]{3}  &  $\mathcal{R}$    & odd  & 1.07771      & 1.03813                                          \\ \hline\hline

                           &                   &even & 1.34176      & 1.15834                                              \\ \cline{3-5}
                           &                   &odd  & 3.84231      & 1.96018                                              \\ \cline{3-5}
                           &  $\mathcal{L}$    &even & 5.85206      & 2.41910                                              \\ \cline{3-5}
                           &                   &odd  & 7.30702      & 2.70315                                            \\ \cline{3-5}
                           &                   &even & 8.79401      & 2.96547                                            \\ \cline{2-5}
\raisebox{2.3ex}[0pt]{5}   &                   &odd  & 1.32835      & 1.15254                                              \\ \cline{3-5}
                           &                   &even & 3.83457      & 1.95821                                              \\ \cline{3-5}
                           &   $\mathcal{R}$   &odd  & 5.84765      & 2.41819                                             \\ \cline{3-5}
                           &                   &even & 7.31179      & 2.70403                                             \\ \cline{3-5}
                           &                   &odd  & 8.80562      & 2.96743                                             \\ \hline

\end{tabular}\\
\caption{The eigenvalue $m^2$ and mass $m$ of the left-and right-handed gravitinos with the odd- and even-parity solutions for the coupling $F(\phi)=\phi^{\alpha}$.
 The parameters are set to $k=1$, $\eta$=1, and $b=1$.}
\label{Tableonephi3}
\end{center}
\end{table*}

\subsection{Localization of the gravitino field in the $f(R)$-thick {branes} with $L=X_1+X_2-V(\phi_1,\phi_2)$}

In the previous subsection, the {$f(R)$-thick branes are} generated by a single canonical scalar field. In this subsection, we will analyze the localization of bulk gravitino in the Bloch-$f(R)$ brane model, where the Lagrangian density of the scalar fields is given by
\begin{eqnarray}
L=-\frac{1}{2}\partial^M\phi\partial_M\phi-\frac{1}{2}\partial^M\xi\partial_M\xi-V(\phi,\xi).
\end{eqnarray}
The scalar fields $\phi$ and $\xi$ interact through the scalar potential $V(\phi,\xi)$. In the following equations, we consider the solution previously given~\cite{Yu:2015wma}:
\begin{subequations}
\begin{eqnarray}
\label{BlochBrane1Phiy}
 \phi(y) \!\!&=&\!\! v \tanh (2dvy),\\
\label{BlochBrane1Chiy}
 \xi(y) \!\!&=&\!\!v \sqrt{\frac{\tilde{b}\!-\!2d}{d}}~\text{sech} (2dvy),\\
\label{BlochBrane1Ay}
 A(y) \!\!&=&\!\! \frac{v^2}{9d}
    \left[ (\tilde{b}\!-\!3d)\tanh^2 (2dvy)
           \!-\!2\tilde{b} \ln \cosh(2dvy) \right],~~~~~~
\end{eqnarray}
\label{BlochBrane1}
\end{subequations}
where $\tilde{b}>2d>0$, and the scalar potential is
\begin{eqnarray}
V(\phi,\xi) &=& \frac12 \left[\left(\tilde{b} v^2-\tilde{b}\phi^2-d\xi^2\right)^2+4d^2 \phi^2 \xi^2 \right]\nonumber\\
&&-\frac43\left(\tilde{b}\phi v^2-\frac13 \tilde{b}\phi^3-d\phi\xi^2\right)^2.
\end{eqnarray}
For certain given values of the parameters $v$ and $\tilde{b}$, the function $f(R)$ could have an analytical expression. For example, when $v=\sqrt{3/2}$ and $\tilde{b}=3d$, we have
\begin{eqnarray}
f(R)&=&R+\frac{2\gamma}{7}\Big[\sqrt{3(R-48 d^2)(R+120 d^2)}\sin\mathcal{Y}(R)\nonumber\\
&&+2\left(R+36 d^2\right)\cos\mathcal{Y}(R)\Big],
\end{eqnarray}
where $\gamma$ is a parameter and $\mathcal{Y}(R)=\sqrt{3}\ln \left(\frac{\sqrt{R-48 d^2}+\sqrt{R+120 d^2}}{2 \sqrt{42}d}\right)$.

Next, we investigate the localization of bulk gravitino with the coupling {function} $F(\phi)=\phi^{p}\xi^{q}$ with $p=1,3,5,\cdots$ and $q$ is any integer. Such coupling was also used to localize the Dirac fermion previously~\cite{Almeida:2009jc,Liu:2009mg,Xie2015}.

\subsubsection{Case I: $F(\phi)=\phi^{p}\xi^{q}$ with $q>0$}

First, we consider the case $F(\phi)=\phi^{p}\xi^{q}$ with $q>0$. For convenience, we let $q=1$. The simplest case is the Yukawa coupling between the two scalar fields and the gravitino, i.e., $-\eta\phi\xi \bar{\Psi}_M [\Gamma^M,~\Gamma^N]\Psi_N$. We also assume without loss of generality that the coupling constant $\eta$ is positive.
The asymptotic behaviors of the potentials (\ref{Vforphi}) in this case are similar to those in the previous subsection. As $z$ (or $y$)$\rightarrow\infty$, both the potentials $V^L$ and $V^R$ vanish, and their values become opposite at $z=0$:
\begin{eqnarray}
V^L(0)=-V^R(0)=2\eta v^3\sqrt{(\tilde{b}-2d){d}},
\end{eqnarray}
which shows that a potential well exists around $z=0$ for $V^R$. Thus, it appears that the left-handed zero mode of the gravitino cannot be localized in the brane, whereas the right-handed zero mode can be localized. However, when substituting the solution of the right-handed zero mode
\begin{eqnarray}
 \chi^{R}_{0}&\propto&\exp\bigg(-\eta\int_{0}^{z}d\bar{z}e^{A(\bar{z})}
       \phi(\bar{z})\xi(\bar{z})\bigg) =
 \exp\bigg( -\eta\int_{0}^{y} d\bar{y}
       \phi(\bar{y})\xi(\bar{y})\bigg)\nonumber \\
  &=&\exp\bigg(\frac{\eta v}{2d}\sqrt{\frac{\tilde{b}-2d}{d}}\text{sech}(2dv\bar{y})|^{y}_{0}\bigg)\nonumber\\
  &\propto& \exp\bigg(\frac{\eta v}{2d}\sqrt{\frac{\tilde{b}-2d}{d}}\text{sech}(2dvy) \bigg)\label{CRight0}
\end{eqnarray}
into the normalization condition (\ref{normalizable condition}), we find the integral
\begin{eqnarray}
 &&\int_{-\infty}^{\infty}(\chi^R_{0}(z))^2 dz
 = \int_{-\infty}^{\infty}(\chi^R_{0}(y))^2 e^{-A(y)} dy \nonumber \\
 &\propto& \int_{-\infty}^{\infty}  \exp\bigg( -A(y)-2\eta\int_{0}^{y}
       \phi(\bar{y})\xi(\bar{y})d\bar{y}  \bigg)dy \nonumber \\
 &=& \mathcal{C}^2\int_{-\infty}^{\infty}\cosh(2dvy)^{\frac{2v^2\tilde{b}}{9d}}\exp\bigg(\frac{2\eta v}{2d}\sqrt{\frac{\tilde{b}-2d}{d}}\text{sech}(2dvy)\nonumber\\
 &&-\frac{v^2}{2d}(\tilde{b}-3d)\tanh^2(2dvy)  \bigg)dy
       \label{fL0CaseI}
\end{eqnarray}
is divergent, which means that the right-handed zero mode cannot be confined to the brane. Although the potential of the right-handed gravitino is a volcanic, no zero mode exists on the brane.
In fact, for any $q>0$ and $p=1,3,5\cdots$, the right-handed zero mode will be a  constant as $y\rightarrow\infty$ because $F(\phi)=\phi^{p}\xi^{q}\varpropto\tanh^p(2dvy)\text{sech}^q(2dvy)\rightarrow 0$. Obviously, this type of a zero mode cannot satisfy the normalization condition (\ref{normalizable condition}). Thus, for any $q>0$, there {exists} no bounded zero mode of the gravitino on the brane (the left-handed zero mode too cannot be localized). Because of the absence of a localized zero mode on the brane, we turn to the case $q<0$.

\begin{figure*}[!htb]
\begin{center}
\subfigure[$V^L$, $p=1$]{\label{fig:VLRCaseII1a}
\includegraphics[width=7cm,height=4.5cm]{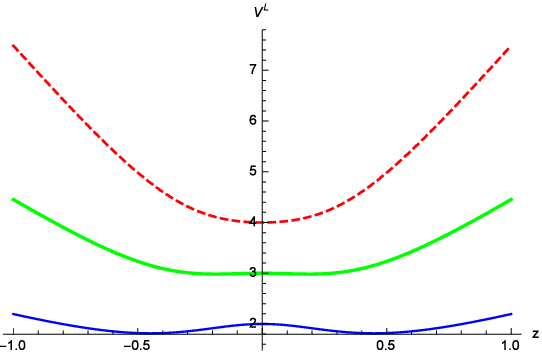}}
\subfigure[$V^R$, $p=1$]{\label{fig:VLRCaseII1b}
\includegraphics[width=7cm,height=4.5cm]{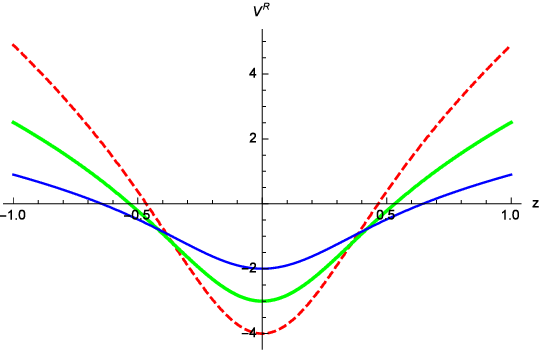}}
\subfigure[$V^L$, $p=3$]{\label{fig:VLRCaseII1c}
\includegraphics[width=7cm,height=4.5cm]{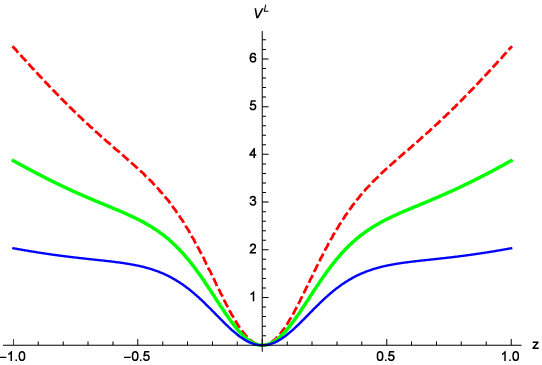}}
\subfigure[$V^R$, $p=3$]{\label{fig:VLRCaseII1d}
\includegraphics[width=7cm,height=4.5cm]{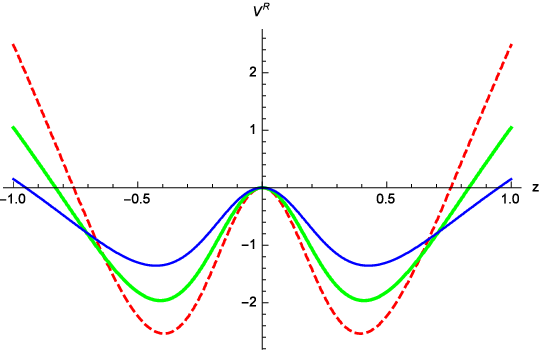}}
\end{center} \vskip -5mm
\caption{Potentials $V^{L}(z)$ and $V^{R}(z)$ for the left- and right-handed gravitinos' KK modes on the $f(R)$-thick {branes} with $F(\phi)=\phi^{p}\xi^{-1}$. Here, $v=d=1$; $\tilde{b}=3$; and the coupling constant $\eta$ is set to $1.0$ (blue thin trace), $1.5$ (green thick trace), and $2.0$ (red dashed trace).}
 \label{fig:VLRCaseIII}
\end{figure*}

\subsubsection{Case II: $F(\phi)=\phi^{p}\xi^{q}$ with $q<0$ (or $q=-1$)}

We let $q=-1$ to represent the case $q<0$ for convenience. The potentials (\ref{Vforphi}) in this case are shown in Fig.~\ref{fig:VLRCaseIII}.
Both the potentials $V^L$ and $V^R$ have infinite wells. For the simplest case $p=1$, both potentials vanish as $z$ (or $y$)$\rightarrow\infty$, and their values become opposite at $z=0$: $V^L(0)=-V^R(0)=\frac{2\eta dv}{\sqrt{(\tilde{b}-2d)/d}}$. The left-handed zero mode yet cannot be localized in the brane because it is divergent as $z\rightarrow\infty$, whereas the right-handed one
\begin{eqnarray}
\chi^{R}_{0}&\propto& \exp\bigg( -\eta\int_{0}^{y} d\bar{y}
       \phi(\bar{y})\xi^{-1}(\bar{y})\bigg)\nonumber\\
  &=& \exp\bigg(-\frac{\eta}{2}\left(\sqrt{{(\tilde{b}-2d)}{d}}~v\right)^{-1}\cosh(2dv\bar{y})|^{y}_{0}\bigg)\nonumber\\
  &\propto& \exp\bigg(-\frac{\eta}{2}\left(\sqrt{{(\tilde{b}-2d)}{d}}~v\right)^{-1}\cosh(2dv\bar{y}) \bigg)
\end{eqnarray}
will vanish as $y\rightarrow\infty$ for any $\eta>0$. Checking whether this right-handed zero mode for any $\eta>0$ can be localized in the brane under the condition \eqref{NormalizationConditionfR0CaseI1Foreta} is easy. Further, for any $q<0$ and $p=1,3,5\cdots$, the right-handed zero mode will be localized. For other $p\geqslant3$, both potentials vanish at $z=0$: $V^L(0)=V^R(0)=0$, and the left-handed potential $V^L$ is always non-negative while $V^R$ have a double well. Therefore, only the right-handed zero mode could be localized in the brane.

There are infinite bounded massive KK modes in this case because both the effective potentials are infinite. Some of our results are listed in Table~\ref{Tabletwophi}. It is clear that the mass spectra of the left- and right-handed gravitinos' massive bounded KK modes are almost the same when their parities are opposite, as shown in the previous section. When $p=1$, the mass of the first bounded state of the left-handed gravitino (or the mass of the first excited state of the right-handed one) increases with the value of $\eta$ because the minima of the left-handed potential $V^L$ increases with $\eta$.
However, the relative width of the effective potentials decreases with the value of $\eta$ and increases with the value of $m^2$.
Thus, the gaps between the bounded states will extend with the growth of $\eta$ and become increasingly narrow as $m^2$ increases. When $p\geq3$, the mass of the first bounded state of the left-handed gravitino still increases with the growth of the $\eta$, despite the minima of the left-handed potential $V^L$ being always zero. Other conclusions are the same as in the case $p=1$.

\begin{table*}[!h]
\begin{center}
\begin{tabular}{||c|c|c|c|c|c||}
 \hline
 ~~~$p$~~~ & ~~~$\eta$~~~  & ~~~$\mathcal{C}$~~~ & ~~~$\mathcal {P}$~~~  & ~~~~~~~$m_n^2$ ~~~~~~~& ~~~~~~~$m_n$ ~~~~~~~  \\
 \hline\hline

                           &    &                   & even& 2.4489      & 1.5649                                           \\
                           \cline{4-6}
                           &    &    $\mathcal{L}$   & odd & 3.6790      & 1.9181                                              \\
                           \cline{4-6}
                           &  &                   & $\vdots$& $\vdots$      & $\vdots$                                              \\
                           \cline{3-6}
                           &  1  &                   & even & 0      & 0                                              \\
                           \cline{4-6}
                           &    &   $\mathcal{R}$     & odd & 2.4490      & 1.5649                                              \\
                           \cline{4-6}
                           &    &                      &  even  & 3.6790      & 1.9181                                             \\
                            \cline{4-6}
                           &    &                   &$\vdots$  & $\vdots$      & $\vdots$                                             \\

                          \cline{2-6} \cline{2-6}

   \raisebox{2.3ex}[0pt]{1}  &    &                   &even & 5.8846      & 2.4258                                              \\
                           \cline{4-6}
                           &    &     $\mathcal{L}$   &odd  & 9.3857      & 3.0636                                             \\
                           \cline{4-6}
                           &    &                   &$\vdots$ & $\vdots$     & $\vdots$                                             \\
                           \cline{3-6}
                           & 2   &                   &even  & 0      & 0                                              \\
                           \cline{4-6}
                           &    &    $\mathcal{R}$   &odd & 5.8849      & 2.4259                                              \\
                           \cline{4-6}
                           &    &                     &even  & 9.3860      & 3.0637                                              \\
                           \cline{4-6}
                           &    &                   &$\vdots$ & $\vdots$ & $\vdots$                                             \\
                           \hline \hline
                           &    &                   & even& 1.8861      & 1.3734                                           \\
                           \cline{4-6}
                           &    &    $\mathcal{L}$   & odd & 3.5178      & 1.8756                                              \\
                           \cline{4-6}
                           &  &                   & $\vdots$& $\vdots$      & $\vdots$                                              \\
                           \cline{3-6}
                           &  1  &                   & even & 0      & 0                                              \\
                           \cline{4-6}
                           &    &      $\mathcal{R}$ & odd& 1.8860      & 1.3733                                              \\
                           \cline{4-6}
                           &    &                     &even  & 3.5177      & 1.8756                                             \\
                            \cline{4-6}
                           &    &                   &$\vdots$  & $\vdots$      & $\vdots$                                             \\

                           \cline{2-6} \cline{2-6}

  \raisebox{2.3ex}[0pt]{3} &    &                   &even & 3.6985      & 1.9232                                              \\
                           \cline{4-6}
                           &    &    $\mathcal{L}$  &odd  & 8.1171      & 2.8491                                             \\
                           \cline{4-6}
                           &    &                   &$\vdots$ & $\vdots$     & $\vdots$                                             \\
                           \cline{3-6}
                           &  2 &                   &even  & 0      & 0                                              \\
                           \cline{4-6}
                           &    &   $\mathcal{R}$   &odd & 3.6981      & 1.9230                                              \\
                           \cline{4-6}
                           &    &                    &even  & 8.1170     & 2.8490                                              \\
                           \cline{4-6}
                           &    &                   &$\vdots$  & $\vdots$     & $\vdots$                                             \\
                           \hline
\end{tabular}\\
\caption{Eigenvalue $m_n^2$ and mass $m_n$ of the bounded left- and right-handed gravitinos' KK modes for the coupling $F(\phi)=\phi^p\xi^{-1}$.
The parameters are set to $v=d=1$ and $\tilde{b}=3$.}
\label{Tabletwophi}
\end{center}
\end{table*}

\section{Discussion and conclusion}\label{secConclusion}

In this manuscript, we investigated the localization and resonant modes of {a} five-dimensional gravitino field on the $f(R)$-thick {branes}, and we considered the Schr\"{o}dinger equations for {the gravitino KK modes} under the gauge condition $\Psi_z=0$. {Similar to the} {five-dimensional free and massless Dirac fermion field}, the zero mode of a free massless five-dimensional gravitino field was localized in a brane only for realizing a compact extra dimension, but its massive KK modes could not realize the localization. Therefore, we introduced the coupling term $-\eta F(\phi)\bar{\Psi}_M[\Gamma^M,~\Gamma^N]\Psi_N$ to investigate the localization of gravitino {on three kinds of $f(R)$-thick branes}. The relative probability method was applied to study the resonances of gravitinos {on these $f(R)$-thick {branes}}. The localization and KK spectra of the five-dimensional gravitino field with the Yukawa coupling term $-\eta F(\phi)\bar{\Psi}_M[\Gamma^M,~\Gamma^N]\Psi_N$ were found to be very similar to {those} of the Dirac fermion, but their chiralities were opposite. This difference may represent the distinction between a five-dimensional Dirac fermion field and gravitino field.

First, we considered the localization of a gravitino on the pure geometric $f(R)$-thick branes, whose Lagrangian density $L(\phi_{i},X_i)$ of the background scalar fields is zero. With the addition of the five-dimensional mass term $\eta F(\phi)=M$, we found that in this system, the KK modes of gravitinos, both the zero mode and massive ones, were unable to localize in the pure geometric $f(R)$-thick branes.

Subsequently, the $f(R)$-thick branes, which are generated by a single canonical background scalar field $\phi$, were considered. We introduced the Yukawa coupling function, $F(\phi)=\phi^\alpha$ with $\alpha=1,~3,~5,~7,~\cdots$ to study the localization of the gravitino field in the $f(R)$-thick brane model. We used {two types of coupling functions} $F(\phi)$, i.e., $\alpha=1$ and $\alpha\geq3$. For the case of $\alpha=1$, localized left- or right-handed zero modes existed on the brane as the coupling parameter $\eta$ satisfied $\eta>\frac{k}{\pi}\sqrt{\frac{2b}{3}}$. Furthermore, for $k=1$ and $b>\frac{1}{2\sqrt{3}}$, we were able to obtain massive resonances of the gravitino on the brane under the condition $\eta>\frac{1}{6}\sqrt{\frac{6+48b+96b^2}{b}}$. The results {indicated} that the left- and right-handed gravitinos had almost the same resonant spectra, but their parities were opposite. With relation (\ref{partiyrelation}), the first resonance of the left-handed gravitino is even and that of the right-handed gravitino is odd. Only the right-handed zero mode of the gravitino was confined on the brane. These results were appropriate for other cases reported in this paper. However, for a five-dimensional Dirac fermion field, only the left-handed zero mode of Dirac fermion was able to localize in the $f(R)$-thick branes, and the first resonance of the left-handed Dirac fermion was odd. The difference in the results of the gravitino and the Dirac fermion field arose from the opposing polarities of $\gamma^5$ in their dynamic equations, which may be a parameter to distinguish the Dirac fermion field and the gravitino field by as they have the same coupling function $F$ and parameter $\eta$. In addition, the number of KK resonant modes for the gravitino {in this braneworld system} increased with an increase in the coupling parameter $\eta$, whereas it decreased with the model parameter $b$. For another case ($\alpha\geq3$), there were no bounded zero modes for both left- and right-handed gravitinos, and the number of KK resonant modes increased with growths in the parameters $b$, $\alpha$, and the coupling parameter $\eta$.

Finally, we focused on the {Bloch-$f(R)$} branes that were generated by two interacting real scalar fields. The coupling function $F(\phi)=\phi^{p}\xi^{q}$ with $p=1,3,5,\cdots$ and $q$ as any integer was considered in this model. For the case $q>0$, there existed no bounded zero modes. For the case $q<0$, the right-handed zero mode was localized in the brane for any $\eta>0$, and there existed infinite bounded massive KK modes for both the left- and right-handed gravitinos because both the effective potentials were infinite potential wells. The gaps between the bounded states extended with the growth of $\eta$ and became increasingly narrow as $m^2$ increased.

Some challenges persisted. As we showed in this paper, the spectra of the KK modes of a bulk gravitino were almost the same as those of a bulk Dirac fermion except for their chiralities. Thus, all the results of the localization of Dirac fermion in branes could be appropriate for the gravitinos by interchanging their chiralities. However, for some kinds of branes, we found the localized KK modes of the Dirac fermion by introducing a new coupling term \cite{Liu:2013kxz,Zhang:2016ksq}. It was not clear whether this coupling term applies to the gravitinos, and it will be our work in the future. In addition, we only considered Minkowski branes in this work. The localization of gravitinos in dS/AdS branes is also interesting.

\Acknowledgements{This work was supported by the National Natural Science Foundation of China (Grant
No. 11647016, No. 11522541, No. 11705106, and No. 11305095), and the Fundamental
Research Funds for the Central Universities (Grant No. lzujbky-2016-k04).}


\end{multicols}
\end{document}